\documentclass{iopjournal}
\usepackage{amsmath,bm,amssymb}

\usepackage[a4paper,left=1in,right=1in,top=2in,bottom=1in,centering]{geometry}

\usepackage{ragged2e}
\usepackage{orcidlink}

\graphicspath{{./}{./Figs/}} 

\newcommand{\const}{\mbox{const}}

\newcommand{\re}{\mbox{Re}}

\newcommand{\ket}[1]{\left| #1 \right\rangle}
\newcommand{\braket}[1]{\left\langle #1 \right\rangle }

\newcommand{\Braket}[2]{\left\langle #1 \middle| #2 \right\rangle}
\newcommand{\BraKet}[3]{\left\langle #1 \middle| #2 \middle| #3 \right\rangle}

\newcommand{\beq}{\begin{eqnarray}}
\newcommand{\eeq}{\end{eqnarray}} 

\newcommand{\hide}[1]{}  %{{\textcolor{red}{[hide]}}}

\newcommand{\TODO}[1]{} %{\bm{\textcolor[rgb]{0.0,0.6,0.2}{#1}}}
 
\newcommand{\Eq}[1]{{\textcolor{blue}{Eq.}}~\!\!(\ref{#1})} 
\newcommand{\Sec}[1]{{\textcolor{blue}{Sec.}}~(\ref{#1})} 
\newcommand{\App}[1]{{\textcolor{blue}{Appendix}}~\ref{#1}} 
\newcommand{\Fig}[1] {{\textcolor{blue}{Fig.}}~\!\!\ref{#1}}

\newcommand{\hrefl}[2]{\href{#2}{(#1)}}

\begin{document}

\title{Quantum measurement of work in mesoscopic systems} 

\author{
Anant Vijay Varma$^{1,2}$\,
\footnote{anant.varma@tuni.fi}
\orcidlink{0000-0002-7610-6317}
and
Doron Cohen$^{1}$\,
\footnote{dcohen@bgu.ac.il}
\orcidlink{0000-0002-3835-3544}
}

\affil{$^{1}$Department of Physics, Ben-Gurion University of the Negev, Beer-Sheva 84105, Israel}

\affil{$^{2}$Computational Physics Laboratory, Tampere University, P.O. Box 600, FI-33014 Tampere, Finland}

\begin{abstract}
\justifying
Heat and work in thermodynamics refer to the measurement of changes in energy content of external bodies (baths and agents). We discuss the implications of quantum mechanics on the possibility to measure work in the mesoscopic context. The agent is a quantum entity (say an oscillator) that is used to drive the system. An inherent limitation is related to back-reaction, leading to a classical-like restriction. We find that in order to resolve fingerprints of interference, an additional quantum uncertainty limitation should be taken into account in the design of the agent. This quantum limitation is fundamental, not merely noise-like blurring.             
\end{abstract}

\justifying

%%%%%%%%%%%%%%%%%%%%%%%%%%%%%%%%%%%%%%%%%%%%%%%%%%%%%%%%%%%%%%%%%%%%%
%%%%%%%%%%%%%%%%%%%%%%%%%%%%%%%%%%%%%%%%%%%%%%%%%%%%%%%%%%%%%%%%%%%%%

\section{Introduction}
\label{s1}

Work ($W$) and heat ($Q$) in thermodynamics refer to energy that is transferred from {\em work agents} and from {\em heat baths} into a system, hence the energy of the system is changed. The first law of thermodynamics implies that the latter equals the sum of the formers, namely, for an $A \leadsto B$ process,
\beq \label{eWQ}
E_B-E_A \ = \ \sum_a W_a + \sum_b Q_b
\eeq
In principle, work and heat can be measured by monitoring energy changes in relatively simple external bodies that exchange energy with the system (a calorimetric approach). A consistent theoretical framework must therefore treat these bodies as dynamical entities. {In particular, work agents are not merely introduced as classical control parameters.}  

Our focus lies on extending these thermodynamic concepts to mesoscopic systems, where fluctuations dominate, and quantum effects become significant -- a framework known as stochastic thermodynamics \cite{tasaki2000jarzynski,kurchan2000quantum,mukamel2003quantum,chernyak2004effect,Huber_2008,talkner2007fluctuation,fusco2014assessing,uzdin2015equivalence,strasberg2022quantum}.
{In this mesoscopic reality work is accessed statistically, yielding a distribution $P(W)$, rather than a definite value,}
and there is special emphasis on non-equilibrium fluctuation theorems
\cite{harris2007fluctuation,esposito2009nonequilibrium,seifert2012stochastic}. 
In particular the Jarzynski and the Crooks relations
refer to the work that is performed on a driven system  
\cite{jarzynski1997nonequilibrium,crooks2000path}. 
Early experimental tests were performed on classical systems, 
e.g. the stretching of an RNA molecule
\cite{liphardt2002equilibrium,Collin_2005}. 
Quantum effects become relevant in experiments on quantum-dot devices  
\cite{saira2012test,koski2013distribution,hofmann2017heat,hofmann2016equilibrium,barker2022experimental,PhysRevLett.131.220405},  
where the time dependent gate voltage is the driving agent.

The actual measurement of {\em changes in energy}, either of the system, or of a bath, or of a work agent, can be carried out using various schemes. 
Naively, one may suggest a two-time measurement scheme 
\cite{talkner2007fluctuation} 
which might be applied for a few-level quantum system  
\cite{smith2018verification}. 
It has variants that involve generalized measurements 
\cite{ancillaPaz,Prasanna_Venkatesh_2014,Lahiri_2021,PhysRevE.108.024126}.
An optional possibility is to measure the associated generating-function, using a Ramsey-like spectroscopic procedure, incorporating an ancilla qubit
\cite{fusco2014assessing,ancilla1,ancilla2,ancilla4}. 

{In the following subsections we introduce the motivation for the present work; clarify conceptually the agent-system setup; explain the experimental advantages; highlight the implied limitations; and specify the specific open questions that have motivated the present study.} 

%%%%%%%%%%%%%%%%%%%%%%%%%%%%%%%%%%%
\subsection{Motivation}
{The above cited literature provides an extensive study of quantum work statistics, treating the work-agent as a non-responsive classical control parameter $X(t)$. As opposed, the present line of study \cite{nft,nfaB,nfa,nfx,Saar} concerns work that is performed by a {\em dynamical} agent. 
We would like to emphasize that the most recent work in this line of study \cite{Saar} by Jarzynski, Deffner and Rahav, titled {\em Fluctuation theorems for autonomous work}, concerns a classical agent, whereas the original motivation that had been stated by Han, Cohen, and Sela \cite{nfa} came from experiments on quantum-dot devices  
\cite{saira2012test,koski2013distribution,hofmann2017heat,hofmann2016equilibrium,barker2022experimental,PhysRevLett.131.220405},
where the time dependent gate voltage is the driving agent. 
In a follow-up paper, Han, Katz, and Sela~\cite{nfx} have further discussed the most direct setting: a superconducting circuit where a transmon, a superconducting qubit, or a qubit array is the driven system, and a microwave resonator or LC mode is the dynamical work agent. The oscillator quadrature supplies the control coordinate \(X\), while the photon number gives access to the energy of the agent.}

%%%%%%%%%%%%%%%%%%%%%%%%%%%%%%%%%%%
\subsection{Work agent}
{We treat the external drive itself as a physical dynamical object and infer work from the energy change of this work agent. The concrete questions addressed below are not about how to define and measure work, but under what design conditions a finite-mass quantum agent reproduces the work statistics of an ideal classically prescribed drive, such that its own back-reaction and quantum uncertainty do not obscure the measurement outcome. To formulate this question quantitatively, we adopt a proper Hamiltonian description of the quantum driving.}
The system is described by Hamiltonian ${ H(\text{sys};X) }$, {where in the common approach $X$ would be regarded a classical control parameter.} For visualization consider \Fig{f0} where $X$ is the position of a piston. 
Within the framework of the work-agent scheme, $X$ is regarded as a dynamical variable. The measurement of {\em work}, within this system-agent framework, has been discussed in the context of the Jarzynski equality in \cite{nft} (see also \cite{nfaB}). Later, emphasizing the experimental perspective, it has been further discussed in \cite{nfa,nfx}. A recent work has extended the formulation of classical fluctuation theorems within this framework \cite{Saar}. It is complementary to schemes involving calorimetric measurements of heat \cite{Qmeas}, i.e. if baths are incorporated  instead of agents.

%%%%%%%%%%%%%%%%%%%%%%%%%%%%%%%%%%%
\subsection{Advantages}
{There are several reasons that motivate the introduction of work-agents and the direct measurement of work: 
{\bf (1)}~Complex circuits might have multi-parametric driving, such that the change in energy is \Eq{eWQ}. One would like to measure separately each $W_a$, and not just the total change of energy.      
{\bf (2)}~Even with single parametric driving, the actual measurement of ${E_B-E_A}$ is in general very difficult, because the energy of a many-body system is not a simple observable, and the realization of a Ramsey-type circuit might be too challenging.
{\bf (3)}~An optional {\em continuous} measurement scheme, that has been used in the mentioned quantum dot experiments, destroys the coherent time-evolution of the system, aka quantum Zeno effect. {We seek non-invasive measurement protocols that preserve the system’s intrinsic dynamics.}} 
%We do not want to perform measurements that change the nature of the dynamics.

%%%%%%%%%%%%%%%%%%%%%%%%%%%%%%%%%%%
\subsection{Limitations}

\begin{figure}
\centering
\includegraphics[width=14cm]{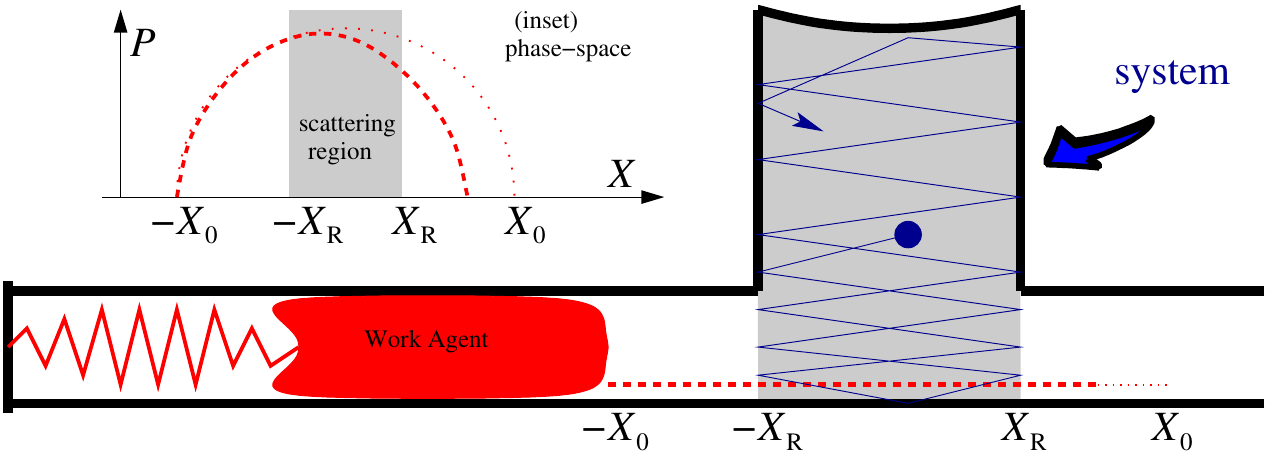} 

\caption{\justifying
{\bf Schematic of the Scattering Process.} 
The system here is a gas particle that is confined in the gray region, say by an optical potential. The agent is a piston connected to a spring. It drives the system as is swings from $-X_0$ towards $X_0$. Due to exchange of energy (work) the velocity of the piston is reduced, and subsequently it does not reach $X_0$. The inset illustrates the trajectory of the agent in phase space (dashed line).  More generally an inelastic collision with a system can either increase or decrease the energy of the agent. We assume that the interaction takes place within the scattering region ${[-X_R,X_R]}$. Outside of this region the energy of the agent is constant.        
}
\label{f0}
\end{figure}

{When the work agent is treated as a dynamical system, two fundamental challenges emerge: (a)~back-reaction and (b)~quantum uncertainty. These can be illustrated by picturing the agent as a piston, see \Fig{f0}}.
An ideal piston would have Hamiltonian $H_{agent}(X,P)=v_0P$ that generates sweep with rate ${\dot{X}}=v_0$. Such a piston would not slow-down if it performed work on a system, and also dispersion in its momentum (due to quantum uncertainty) will have no adverse consequences. 
{However in practice the agent is not ideal. The agent is an object that has finite mass~$M$, such that $v=(1/M)P$, and accordingly $v_0$ is not sharply defined due to the uncertainty in momentum, and it will be affected due to back-reaction.}

%%%%%%%%%%%%%%%%%%%%%%%%%%%%%%%%%%%
\subsection{Open questions}

{A fundamental question is how to prescribe the design of a “good” work agent -- one that suppresses back-reaction and quantum uncertainty, thereby approaching the performance of an idealized agent. This issue was addressed in \cite{nfa}, albeit within a narrow scope and without a definitive resolution.}
{The back-reaction issue (aka Born–Oppenheimer condition) is straightforward, as it follows from classical reasoning. In contrast, the role of quantum uncertainty is far less obvious:} 
{Is it just a matter of blurring the outcome of the measurement? If so, maybe it is not a fundamental limitation? Maybe there is a way to overcome this limitation, say by adopting a super-resolution technique? 
An emergent subtlety arises if the agent is an LC circuit that swings with amplitude that involves $n_{ph}$ photons \cite{nfa}. To minimize back-reaction large $n_{ph}$ is favored, while for minimizing uncertainty, contrary to naive intuition, the opposite is desired. The implied dilemma is phrased very clearly in a subsequent recent publication by Han, Katz and Sela \cite{nfx} while discussing the experimental aspect: ``The energy measurement of a coherent state suffers from an uncertainty $\sqrt{n_{ph}}$. Minimizing this uncertainty, and also satisfying the Born-Oppenheimer condition, allows access to the work distribution function only in a limited parameter regime".}       
  
{The focus here is on experiments where quantum mechanics is probed. The fingerprints of quantum interference are fluctuations in $P(W)$, analogous to Universal Conductance Fluctuations. The "strength" of those fluctuations is expected to be controlled by ``small parameter" that is related to the design of the work agent. The practical question that arises is whether the fingerprints of interference are endangered by ``bad" design of the work agent, what is the ``small parameter", and is it related to the quantum uncertainty issue. This is the central question we aim to address here.}

%%%%%%%%%%%%%%%%%%%%%%%%%%%%%%%%%%%%%%%%%%%%%%%%%

%%%%%%%%%%%%%%%%%%%%%%%%%%%%%%%%%%%
\subsection{Outline}

{The system-agent setup is defined in \Sec{sSetup}, where the definitions of the work distribution function $P(W)$ and its idealized version $P^{cl}(W)$ are presented. \App{sIdeal} provides the formal proof that an ideal agent does not require a classical reality.
In \App{s9} we provide a concise review of the two point measurement scheme, and for sake of completeness also discuss the option to use a Ramsey-like procedure for the measurement of {\rm work}.} 
 
In order to have a clear formulation of the agent design problem we present a scattering perspective for the system-agent interaction. Consequently, in \Sec{s3}, all the design considerations are expressed as conditions that concern the control over the sweep velocity $v_0$. We point out that quantum uncertainty is relevant for two conceptually different conditions. It turns out later that the additional condition that we highlight is relevant for ensuring the manifestation of quantum interference.  

Specifically we discuss the design considerations for an oscillator that serves as an agent in \Sec{s4}, leading to the design diagram of \Fig{f1}. A critical discussion follows in \Sec{s5}, and motivates the introduction of a model system in \Sec{s6}.  Numerical demonstrations are presented in \Sec{s6s}, and analyzed in \Sec{s7}. The quantum condition for resolving interference is worked out in \Sec{s8}. Finally we conclude in \Sec{s10}.

\clearpage  
%%%%%%%%%%%%%%%%%%%%%%%%%%%%%%%%%%%%%%%%%%%%%%%%%%
\section{{The system-agent setup}}
\label{sSetup}

The Hamiltonian of the system and the agent can be written schematically as 
\beq
H_{total}=H(sys;X)+ H_{agent}(X,P)
\eeq
The instantaneous eigenstates of $H(sys;X)$ for a given $X$ are 
$\ket{\nu(X)}$ with eigen-energies $E_{\nu}(X)$. 
The Born-Oppenheimer basis of the agent-system complex consists of the states ${\ket{X,\nu(X)}}$. 
Using scattering theory terminology we refer to $\nu$ as channel index. 
The scattering scenario is caricatured in \Fig{f0}.   
The agent itself is chosen below to be either idealized, or a piston or an oscillator: 
\beq \label{eIWA}
H_{agent}^{(idealized)} &=& \ v_0 P 
\\  \label{eHP}
H_{agent}^{(piston)} &=& \ \frac{1}{2M} P^2 
\\ \label{eHO}
H_{agent}^{(osc)} \ &=& \ \frac{\omega}{2} \left[ \left(\ell \hat{P} \right)^2 + \left( \frac{\hat{X}}{\ell} \right)^2 \right]
\eeq
The scenario of interest assumes that the agent sweeps through the system with velocity $\sim v_0$. The interaction takes place within a scattering region ${[-X_R,X_R]}$. Outside of this region the energy of the agent is constant. We shall use $\varepsilon$ and $E$ to indicate the energies of the agent and of the system respectively. The subscripts $A$ and $B$ will indicate the values before and after the interaction.        
One would like to measure the work, i.e. the energy that is transferred from the agent to the system: 
\beq
W \ \equiv \ -(\varepsilon_B -\varepsilon_A) \ = \ E_B-E_A  
\eeq      
In this equation the first equality is the definition of work,
while the second equality is implied by conservation of energy, 
aka the first law of Thermodynamics. In a more complicated setups one might have several agents and several baths. Needless to mention that in general the work can be either positive or negative.

%%%%%%%%%%%%%%%%%%%%%%%%%%%%%%%%%%%%%%%%%%%
\subsection{{An idealized agent}}

An idealized classical driving assumes that $X$ can be regarded as a classical control parameter that is changed such that $\dot{X}=v_0$ during the interaction with the system. If the agent is an oscillator, it is assume that 
\beq \label{eXcl}
X^{cl}(t)=-X_0 \cos(\omega t), 
\ \ \ \ t \in \left[0,\frac{\pi}{\omega}\right]
\eeq
where the frequency $\omega$ and the amplitude $X_0$ are adjusted such that the sweep velocity is ${v_0 = X_0\omega }$ during the interaction with the system at ${X \sim 0}$. 
For an idealized driving we define the spectral function
\beq \label{ePcl}
P^{(cl)}(W) = \sum_{\nu} 
\left| \BraKet{\nu}{U}{\nu_0}  \right|^2  
\delta(W{-}(E^{(B)}_{\nu}{-}E^{(A)}_{\nu_0})) 
\eeq
where $U$ is the evolution operator of the sweep process.
Here $E^{(A,B)}_{\nu}$ are the eigen-energies of the Hamiltonians of the uncoupled system, namely,  
\beq \label{eHAB}
H^{(A,B)} \ \ = \ \ H(sys; X_{A,B}) 
\eeq
where $X_{A,B}$ are arbitrary locations before and after the scattering region. The spectral function $P^{(cl)}(W)$ characterizes the statistics of the change in the energy of the system, without taking into account the dynamical nature of the agent.

We would like to regard the agent as a dynamical entity. A preliminary question that arises is whether in a quantum context it is possible to define a ideal quantum agent that produces the same distribution $P^{(cl)}(W)$. The answer is positive. We can rigorously prove (see \App{sIdeal}) that an ${M=\infty}$ agent, described by \Eq{eIWA}, is equivalent to parametric classical driving.

%%%%%%%%%%%%%%%%%%%%%%%%%%%%%%%%%%%%%%%%%%%
\subsection{{A realistic agent}}

The question arises what are the design considerations for a non-ideal realistic quantum agent that has a finite mass~$M$. In order to make further progress we have to introduce a proper definition for a spectral function that characterizes the work, i.e. the statistics of the change in the energy of the agent.  Let us assume that the agent is an oscillator with eigen-energies ${\varepsilon_n = \omega n }$, where ${n=0,1,2,\cdots}$ is the excitation index (to which we refer later as the number of photons). 
Formally the tendency would be to define the spectral function 
\beq \label{ePw}
P^{qm}(w) = \sum_{n,\nu} 
\left| \BraKet{n,\nu}{U}{n_0,\nu_0}  \right|^2  
\delta(w{-}(\varepsilon_{n}{-}\varepsilon_{n_0})) 
\eeq   
where $\nu_0$ is the initial energy-state of the system, 
and $n_0$ is the initial energy-state of the agent. 
{Here $U$ is the evolution that is generated by $H_{total}$.}
Note that ${W=-w}$. 
While this definition might be useful for a {\em different} type of experiment, it is not appropriate here because we have an interest in a classical-like driving scenario that is described by \Fig{f0}. 
The initial state of the agent has to be a localized wavepacket, preferably a coherent state, else the notion of {\em sweep} is meaningless.

We use the notation $\alpha_0$ to indicate the initial coherent state preparation of the agent, such that ${X \sim -X_0}$ and ${P\sim 0}$. We define {\em preparation Hamiltonian} of the agent as follows:  
\beq \label{eHPrep}
H^{(A)}_{agent} \ = \ \frac{\omega}{2} \left[ \left(\ell \hat{P} \right)^2 + \left( \frac{\hat{X}+X_0}{\ell} \right)^2 \right] 
\eeq
By definition the coherent state $\alpha_0$ is the ground state of this {\em shifted} Hamiltonian. It is a minimal wavepacket that has uncertainty $\ell$ in its position.  Now it is possible to re-define the {\em initial} energy in \Eq{ePw}, relative to the ground state, as 
\beq
\varepsilon_{n_0}^{(A)} \ \ \equiv \ \ \frac{\omega}{2}\left(\frac{X_0}{\ell}\right)^2 + n_0^{(A)}\omega  
\eeq
where $n_0^{(A)}$ is the initial excitation relative to the preparation Hamiltonian. By default the ideal preparation would be with ${n_0^{(A)}=0}$. Consequently, the practical definition of the work distribution function takes the form 
\beq \label{ePW}
P(w) = \sum_{n} 
P_n \, \delta(w{-}(\varepsilon_{n}{-}\varepsilon_{n_0}^{(A)})) 
\eeq   
where ${P_n=\sum_{\nu} \left| \BraKet{n,\nu}{U}{\alpha_0,\nu_0} \right|^2}$ is the probability to find the agent in the $n$-th excitation level. Here ${U=\exp[-iH_{total}t]}$. We remind that the common sign convention for work in the NFT literature is ${W=-w}$. 
We briefly review how changes in energy can be actually measured in \App{s9}.

%%%%%%%%%%%%%%%%%%%%%%%%%%%%%%%%%%%%%%%%%%%%%%%%%%
\section{The design of the agent}
\label{s3}

The design of the work agent requires the specification of the desired driving rate $v_0$, and the desired accuracy $\delta v_0$. Additionally we have to know in advance the typical work that has to be measured. For simplicity we assume that there is single energy scale $W_0$ that characterizes the expected work distribution $P(W)$. 

\begin{figure}[t!]
\centering

\includegraphics[width=8cm]{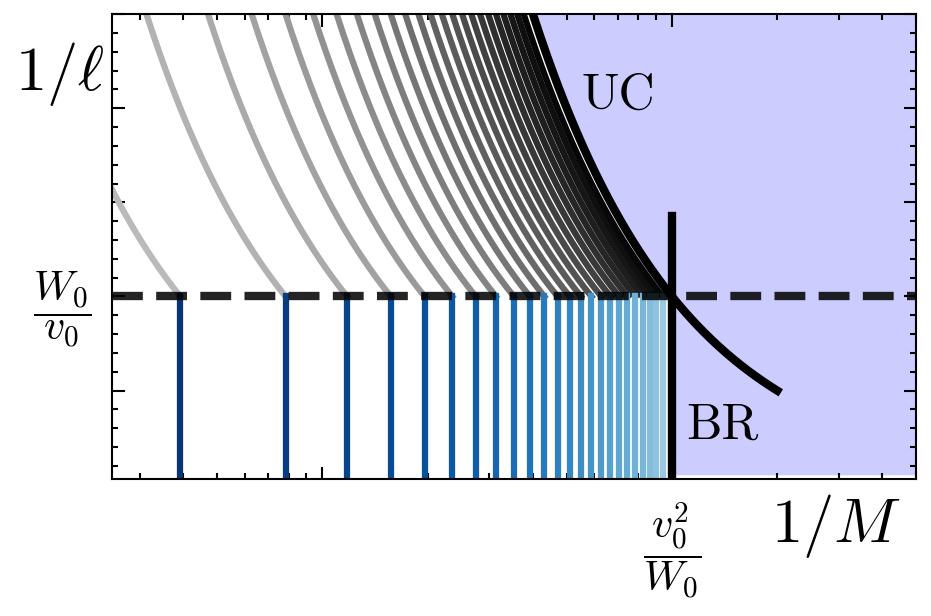}
\\ 

\includegraphics[width=8.5cm]{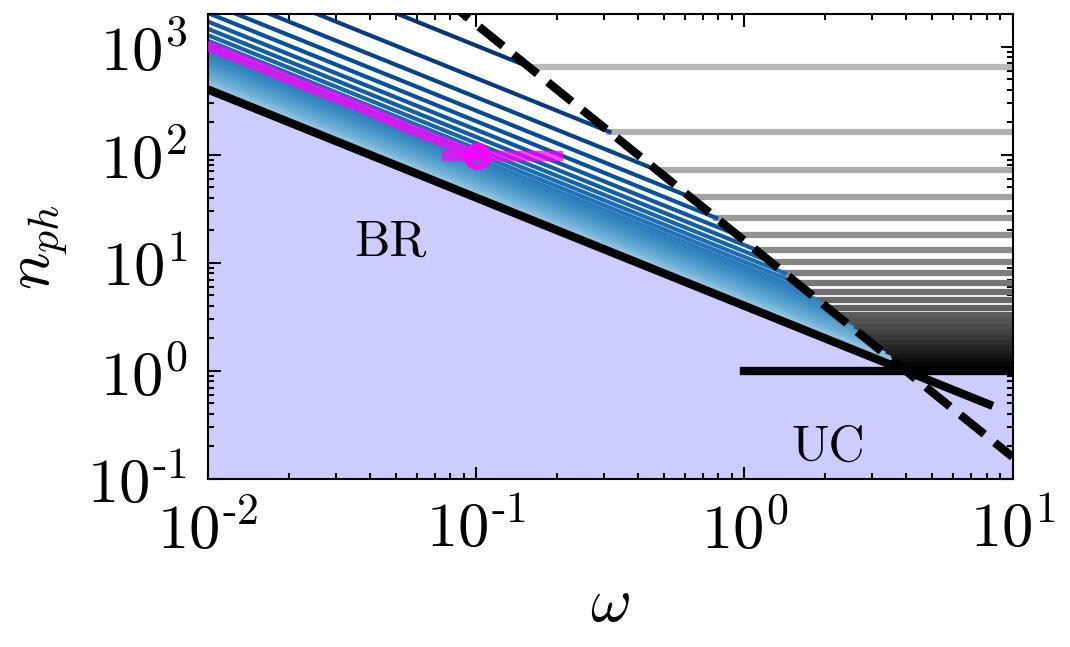}

\caption{\justifying
{\bf Work agent design diagram.} 
Given the design goal parameters of the measurement ${(W_0,v_0,\delta v_0)}$, the various constraints on the free parameters of the agent are indicated. Upper and lower panels are respectively for the ${(M,\ell)}$ of the piston and for the ${(\omega,n_{ph})}$ of the oscillator that are used as work agents. The dashed line is the ${\delta v_{uc} \sim \delta v_{br}}$ border. The bold solid lines "UC" and "BR" indicates the ${\delta v_{uc} < \delta v_0}$ and the ${\delta v_{br} < \delta v_0}$ borders respectively {for ${\delta v_0 = v_0}$, while the other solid lines are the same borders for smaller values of $\delta v_0$.} In the lower panel these are ${\ell \omega {=} \const}$ and ${\ell^2 \omega {=} \const}$ contours along which either $\delta v_{uc}$ or $\delta v_{br}$ are constant, respectively. Magenta dot and lines indicate numerical availability of numerical results in subsequent figures.}
\label{f1}
\end{figure}

If the agent is a piston, whose Hamiltonian is \Eq{eHP}, the free design parameters are ${(M,\ell)}$. They determine the initial momentum ${P_0=Mv_0}$, and its uncertainty ${\delta P_0 \approx 1/\ell}$. From that we get the initial uncertainty  $\delta v_{uc}$ of the sweep rate, and the generated dispersion $\delta v_{br}$ due to back-reaction. Namely,  
\beq
\delta v_{uc} &=& (1/M)[1/\ell]    \\ 
\delta v_{br} &=& (1/M) [W_0/v_0] 
\eeq
The agent design requirements are 
\beq \label{eUCp}
\delta v_{uc} < \delta v_0 
\ \ \ &\leftrightsquigarrow& \ \ \  
\left(\frac{1}{M}\right)\left(\frac{1}{\ell}\right) < \delta v_0
\\ \label{eBRp}
\delta v_{br} < \delta v_0 
\ \ \ &\leftrightsquigarrow& \ \ \  
\left(\frac{1}{M}\right)  < \frac{v_0\delta v_0}{W_0}  
\eeq
Additionally, in order to resolve changes in the energy of the agent, it makes sense to require that ${\delta E = v_0 \delta P}$ would be much smaller compared with $W_0$. This extra requirement can be written optionally as 
\beq \label{eUCr}
\delta v_{uc} < \delta v_{br} 
\ \ \ &\leftrightsquigarrow& \ \ \  
\frac{1}{\ell} < \frac{W_0}{v_0}
\eeq
An $(M,\ell)$ design diagram is presented in \Fig{f1}. Note that one may speculate that super-resolution techniques can be used to relax the requirement of \Eq{eUCr}, as opposed to the more fundamental requirements of \Eq{eUCp} and \Eq{eBRp}.

%%%%%%%%%%%%%%%%%%%%%%%%%%%%%%%%%%%

\clearpage 
%%%%%%%%%%%%%%%%%%%%%%%%%%%%%%%%%%%%%%%%%%%%%%%%%%
\section{Oscillator as an agent}
\label{s4}

In an experiment, it is more natural to use an oscillator as an agent, typically an LC circuit \cite{nfa,nfx}.  The Hamiltonian is written conveniently as in \Eq{eHO}, 
or optionally as ${H_{agent} = \omega \bm{b}^{\dag}\bm{b} }$. The occupation coordinate is ${ \bm{n}=\bm{b}^{\dag}\bm{b} }$ with eigenvalues ${n=0,1,2,\cdots}$. 
The oscillator is prepared in a coherent state at ${X=-X_0}$, and then swings to ${X \sim X_0}$. If one could neglect back-reaction, and regard the agent as a classical entity, the driving would be \Eq{eXcl}.  
Specifically the system Hamiltonian is characterized by a scattering region ${|X| < X_R}$ that is possibly much smaller than the sweep amplitude $X_0$. Effectively,  the oscillator agent is not much different from a piston agent.

In the quantum reality, we have $\ell$ as an extra free parameter.  It is implied that the mass of oscillator is ${M=1/(\ell^2 \omega)}$, and that the initial momentum uncertainty during the interaction is ${\delta P \approx 1/\ell}$, as in the case of the piston agent.  The number of photons that are required for launching the agent (disregarding factor of order unity) is 
\beq
n_{ph} \ = \ \left(\frac{X_0}{\ell}\right)^2 \ = \ \left(\frac{v_0}{\omega\ell}\right)^2
\eeq
Rather than using ${(\omega, \ell)}$ as free design parameters, it is more illuminating to use ${(\omega, n_{ph})}$ as free parameters. Accordingly,
\beq
\delta v_{un} &=&  \ell \omega  
\ = \ \frac{1}{\sqrt{n_{ph}}} v_0
\\ 
\delta v_{br} &=& \left[\frac{W_0}{v_0}\right] \ell^2 \omega  
\ = \ \frac{1}{n_{ph}} \left(\frac{W_0}{\omega}\right)
\eeq
The dimensionless parameter $n_{ph}$ serves in some sense as a scaled inverse Planck constant. Note that $\delta v_{br}$ is a classical quantity because $\hbar\omega$ scales as classical energy.   
One concludes that the agent design requirements are 
\beq \label{eUC}
\delta v_{uc} < \delta v_0 
\ \ \ &\leftrightsquigarrow& \ \ \  
n_{ph} > \left(\frac{v_0}{\delta v_0}\right)^2
\\ \label{eBR}
\delta v_{br} < \delta v_0 
\ \ \ &\leftrightsquigarrow& \ \ \  
n_{ph} >  \left(\frac{v_0}{\delta v_0}\right) \left(\frac{W_0}{\omega}\right)
\\ \label{eRes}
\delta v_{uc} < \delta v_{br} 
\ \ \ &\leftrightsquigarrow& \ \ \  
n_{ph} < \left(\frac{W_0}{\omega}\right)^2
\eeq
These borders are illustrated in \Fig{f1} in one-to-one correspondence with the piston agent diagram.

%%%%%%%%%%%%%%%%%%%%%%%%%%%%%%%%%%%%%%%%%%%%%%%%%%
\section{Critical discussion}
\label{s5}

In \cite{nfa}, the above \Eq{eBR} and \Eq{eRes} were written as strong inequalities, namely, 
\beq
(W_0/\omega) \ll n_{ph}  \ll  (W_0/\omega)^2 
\eeq
It is implicit that the oscillator should swing with a frequency ${\omega \ll W_0}$.  
This implicit assumption is rather obvious, because what we measure in practice is the occupation $n$ of an harmonic mode, and from that we get a distribution $P(W)$ that is discretized in units of $\omega$. Surely we want a resolution that is better than $W_0$.    
To avoid confusion, it should be clear that $W_0$ refers here to the energy scale that should be resolved at the time of the measurements, and not e.g. to energy scales that characterize e.g. Landau-Zener transitions during the system-agent interaction.

Let us speculate that super-resolution techniques can be used in order to extract $P(W)$ from the measurement of agent energy even if \Eq{eRes} is not satisfied, meaning that the blurring is eliminated via an appropriate de-convolution.  In such case the uncertainty considerations become redundant for ${\omega \ll W_0  }$. The following discussion reveals that this conclusion is misleading.  It is based on a wrong assumption that the same $\delta v_0$ is required in both the BR and the UC conditions.

Thus, in order to refine our reasoning, we have to clarify how the design parameter $\delta v_0$ is determined. In \cite{nfa} the subtext was that the required $\delta v_0$ should be simply a reasonable fraction of $v_0$. This leads to the  rather innocent requirement  ${n_{ph} \gg 1}$.   
But in fact there is a much stronger requirement. The probability distribution $P(W)$ depends on $v_0$, and this dependence might be sensitive to quantum interference. If we want to resolve the quantum fingerprints, we have to ensure high control over $v_0$, which means a very small $\delta v_0$.  We will discuss the actual determination of $\delta v_0$ in \Sec{s8}. 
But first we have to introduce our model system.

%%%%%%%%%%%%%%%%%%%%%%%%%%%%%%%%%%%%%%%%%%%%%%%%%%
\section{Model system}
\label{s6}

We consider a 2-site (dimer) or 3-site (trimer) Bose Hubbard system, where the parameter~$X$ controls the potential difference between the first and last sites. The sweep protocol is known as many-body Landau-Zener (LZ) transition or as Rapid Adiabatic Passage (RAP) respectively. The system Hamiltonian is 
\beq
&& H(sys;X) \ = \   
\frac{U}{2} \sum_{j=1,L} \bm{n}_{j}^{2} 
- \frac{K}{2} \sum_{j<L} (\bm{a}^{\dagger}_{j+1} \bm{a}_{j} + h.c.)
\nonumber \\
&& +\dfrac{X_c}{2} \tanh\left(\frac{X{-}X_{a}}{X_c}\right) \bm{n}_{1}
-\dfrac{X_c}{2} \tanh\left(\frac{X{+}X_{a}}{X_c}\right) \bm{n}_{L} 
\ \ \ \ \ \ 
\label{E1}
\eeq
For the dimer $X_{a}=0$, while for the trimer it is a free parameter (see below). 
The second line in this equation implies that the interaction takes place within a ``scattering region" of limited range. Outside of the interaction region the motion of the agent is ``free", uncoupled from the system. 
During the interaction the potential of the first site rises from $-X_c/2$ to $+X_c/2$, while the potential of the last site drops from $+X_c/2$ to $-X_c/2$.  For the trimer, the extra parameter $X_{a}$ allows a delay between the first and the last level crossings. In practice we chose $X_{a}=X_{c}/4$, reflecting that from numerical perspective the range of the interaction is ${X_R \sim (1/4)X_c + X_a}$. 

\ \\ 

%%%%%%%%%%%%%%%%%%%%%%%%%%%%%%%%%%%%%%%%%%%%%%%%%%%
{\em Entanglement.-- }
%\label{s2}
%
The initial state of the system is ${\nu=\nu_0}$.
The later state of the system is in general an entangled superposition with normalized wavepackets $\psi^{(\nu)}(X;t)$ that travel in the different channels. This superposition can be written at a given time as follows:    
\beq
\ket{\Psi(t)} 
\ \ &=& \ \ \sum_{X,\nu} \Psi_{\nu}(X) \ket{X,\nu(X)} \\
\ \ &=& \ \ \sum_{\nu} \sqrt{p_{\nu}} \ |\psi^{(\nu)}\rangle \otimes \ket{\nu}
%\ket{\Psi(t)}_{qm} \ \ &=& \ \ \sum_{\nu} \sqrt{p_{\nu}} \ |\psi^{(\nu)}\rangle \otimes \ket{E_{\nu}(X^{qc}(t)}
\eeq
In the first line the wavefunction is expanded in the Born-Oppenheimer basis. In the second line we regard $\nu$ as a dummy channel index, which allows to formally define a normalized wavefunction $\psi^{\nu}$ for each channel.

%%%%%%%%%%%%%%%%%%%%%%%%%%%%%%%%%%%%%%%%%%%%%%%%%%%%%%%
\begin{figure}
%\raggedright

\centering

\includegraphics[height=4cm]{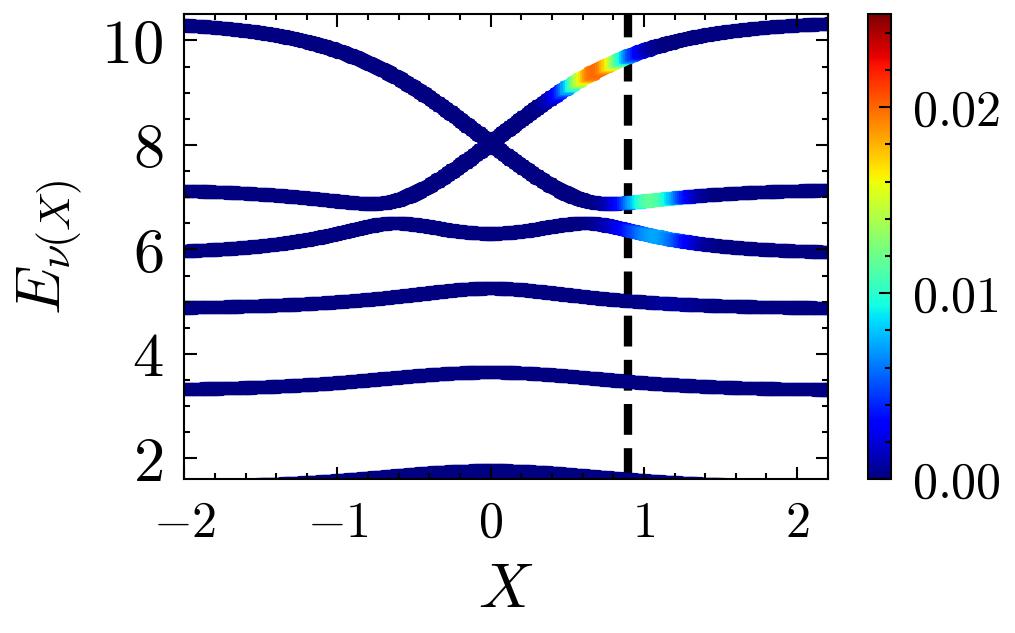} 
\includegraphics[height=4cm]{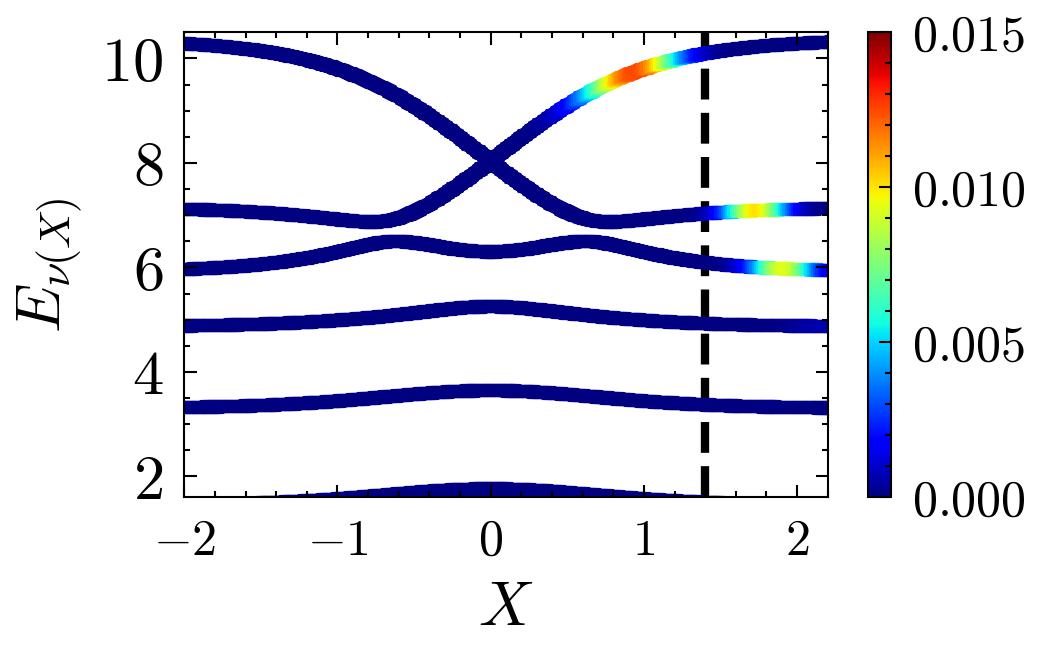}

\includegraphics[height=4cm]{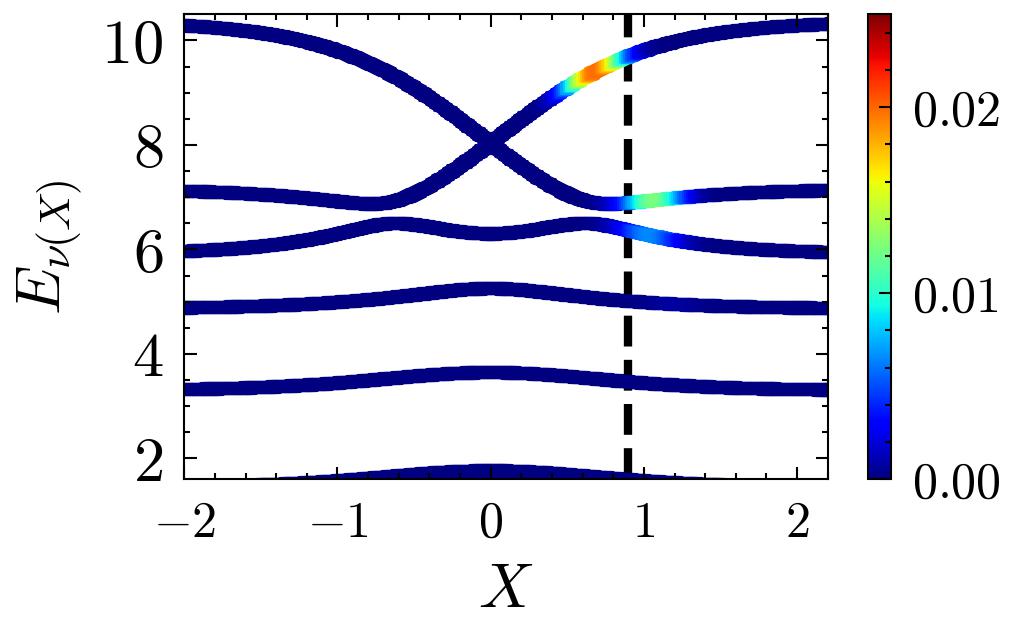}  
\includegraphics[height=4cm]{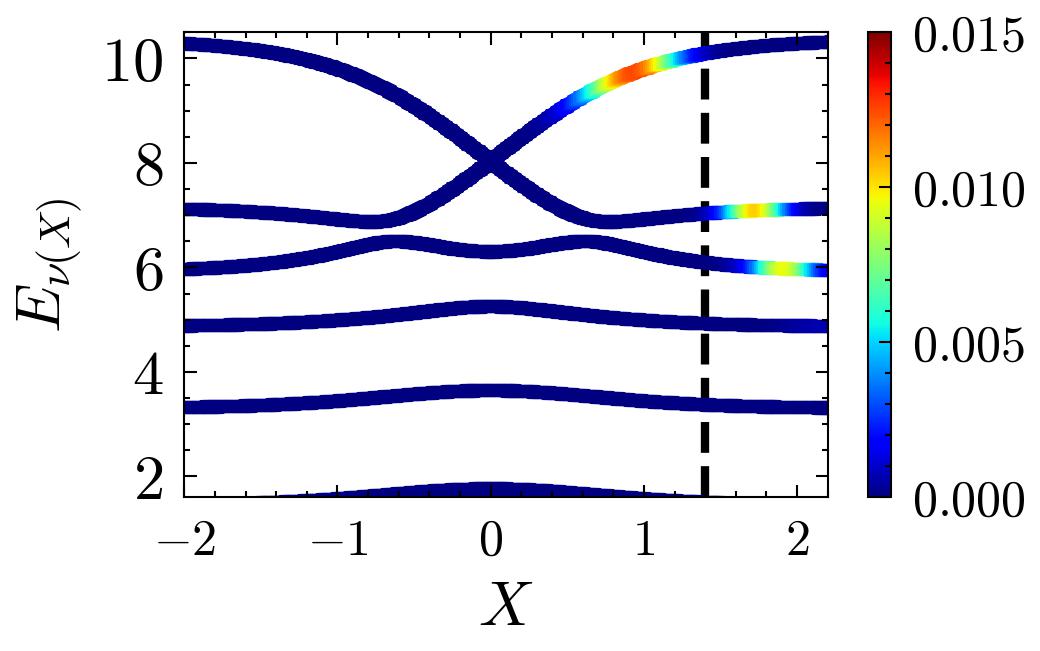} 

\caption{\justifying
{\bf Quasiclassical representation.}
Representative images of the probability distribution 
${P(X,\nu(X))}$, indicated by color,  
when ${X^{cl}(t)=1.2}$ (left panels) 
and when ${X^{cl}(t)=2}$ (right panels), 
for the simulation that is later discussed in \Fig{f2}b. 
The average position ${X^{qc} \equiv \braket{X}_t}$ is indicated by vertical dashed lines. In the lower panels the calculation has been done using the approximated transformation matrix as discussed in \App{sBasis}.
}
\label{f2p}
\end{figure}

{All the subsequent calculations are based on straightforward calculation of the probability distribution}
\beq
P(X,\nu) 
\ = \ \left| \Psi_{\nu}(X) \right|^2
\ = \ \left|\BraKet{X,\nu}{U}{\alpha_0,\nu_0}\right|^2
\eeq
{which is illustrated in \Fig{f2p}.} 
In particular the work distribution function is determined by \Eq{ePW} with 
\beq
P_n \ = \ \sum_{\nu} p_{\nu} \left|\Braket{n}{\psi^{(\nu)}}\right|^2 
\eeq

The quasi-classical motion of the agent in channel~$\nu$ 
is described by ${X^{\nu}(t)=\BraKet{\psi^{(\nu)}}{X}{\psi^{(\nu)}} }$, and the global average is 
\beq \label{eXqc}
X^{qc}(t) \ = \ {\braket{X}_t}
\ = \ \sum_{\nu} p_{\nu}(t) X^{\nu}(t)
\eeq

{The calculation of the $\Psi_{\nu}(X)$ representation from the standard representation $\Braket{x,n_j}{\Psi}$ of the numerical code is explained in \App{sBasis}. It requires $N_{osc}$ diagonalizations of the Bose-Hubbard Hamiltonian, where $N_{osc}$ is the Hilbert space dimension of the agent. In each diagonalization one obtains a transfer matrix $\Braket{n_j}{\nu(X)}$.  In practice, if the wavefunction $\Psi$ is well localized in $X$, one can use as an approximation a single matrix $\Braket{n_j}{\nu(X^{qc})}$, hence the numerical cost of a simulation is reduced. This procedure is rather accurate even in the challenging example of \Fig{f2p}.}   

There are two extreme approximations that can be used in order to ``factorize" the entangled state. In one extreme one assumes that all the wavepacket ${ \psi^{(\nu)} }$ are the same up to a phase, hence the system remains coherent, described by   
${ \sum_{\nu} \sqrt{p_{\nu}} e^{i\varphi_{\nu}} \ \ket{\nu} }$.
The other extreme possibility is that wavepackets ${ \psi^{(\nu)} }$ are non-overlapping, though possibly have approximately the same position, hence the state of the system can be described as a non-coherent mixture of $\nu$ states with probabilities $p_{\nu}$. % 
In both limiting cases, the system dynamics is governed by $X^{qc}(t)$, still the nature of the dynamics is different: either coherent or stochastic.

%%%%%%%%%%%%%%%%%%%%%%%%%%%%%%%%%%%%%%%%%%%%%%%%%%%%%%%
\begin{figure}
%\raggedright

{\bf (a)} \hspace{5cm}  {\bf (b)} \hspace{5cm}  {\bf (c)}  \\

\includegraphics[height=3.5cm]{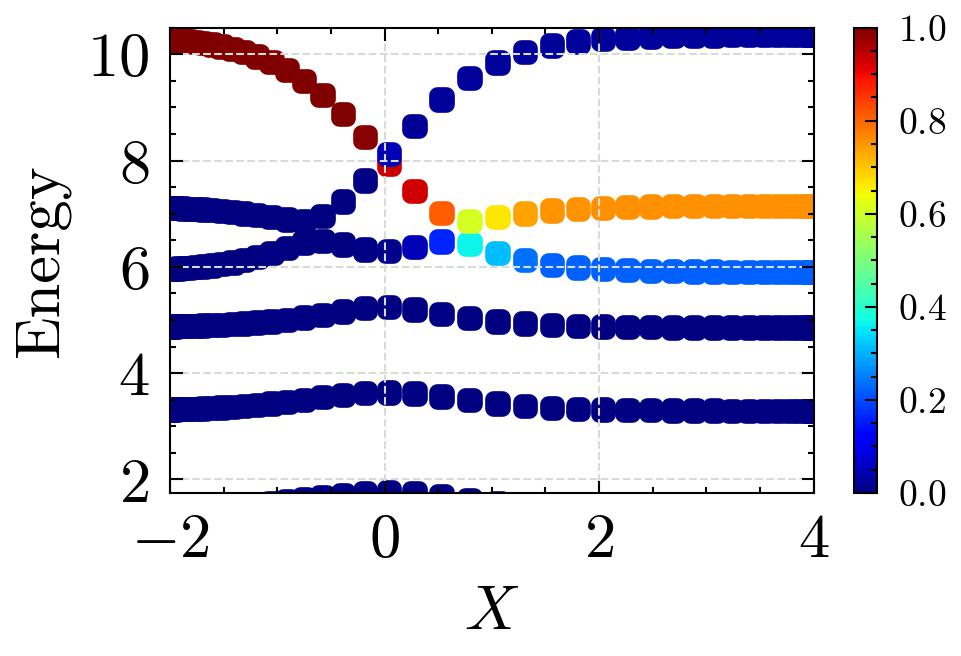} 
\includegraphics[height=3.5cm]{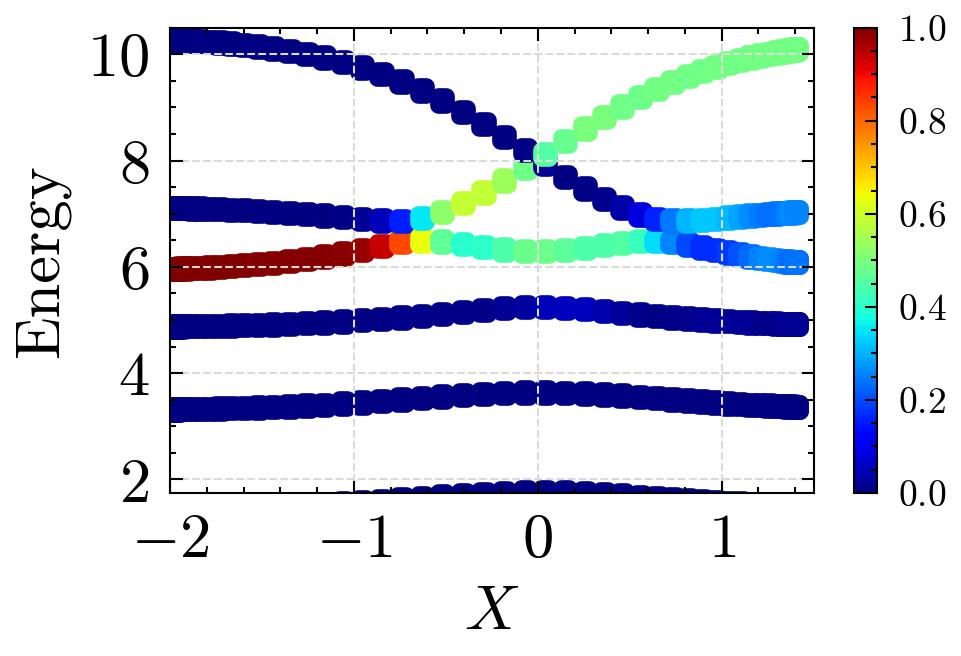} 
\includegraphics[height=3.5cm]{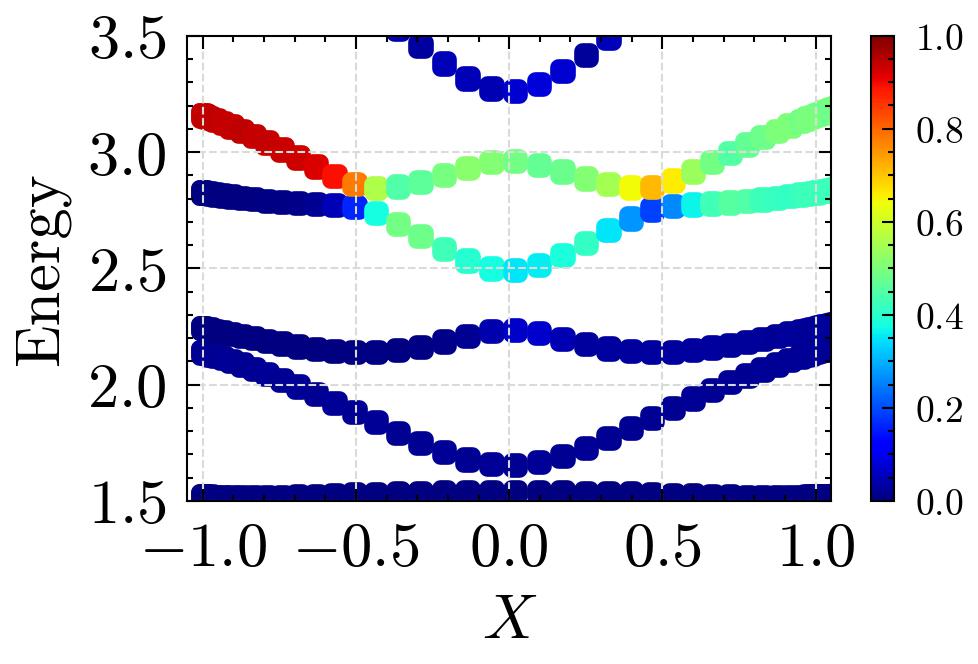} 

\hspace*{1mm}
\includegraphics[height=3cm]{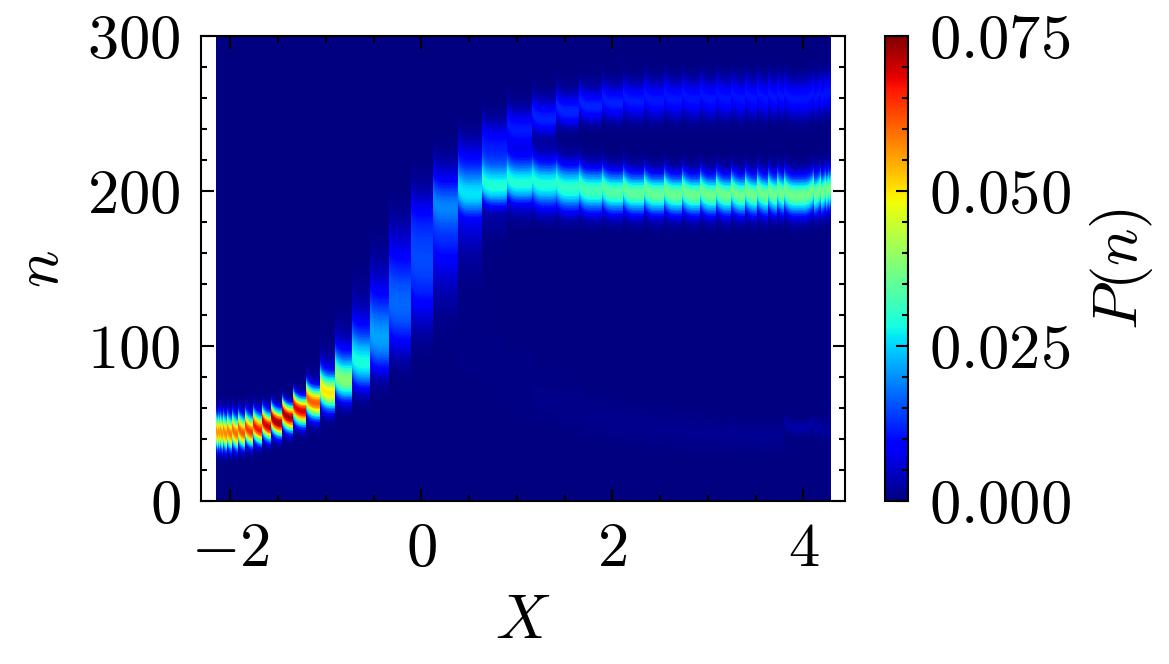} 
\includegraphics[height=3cm]{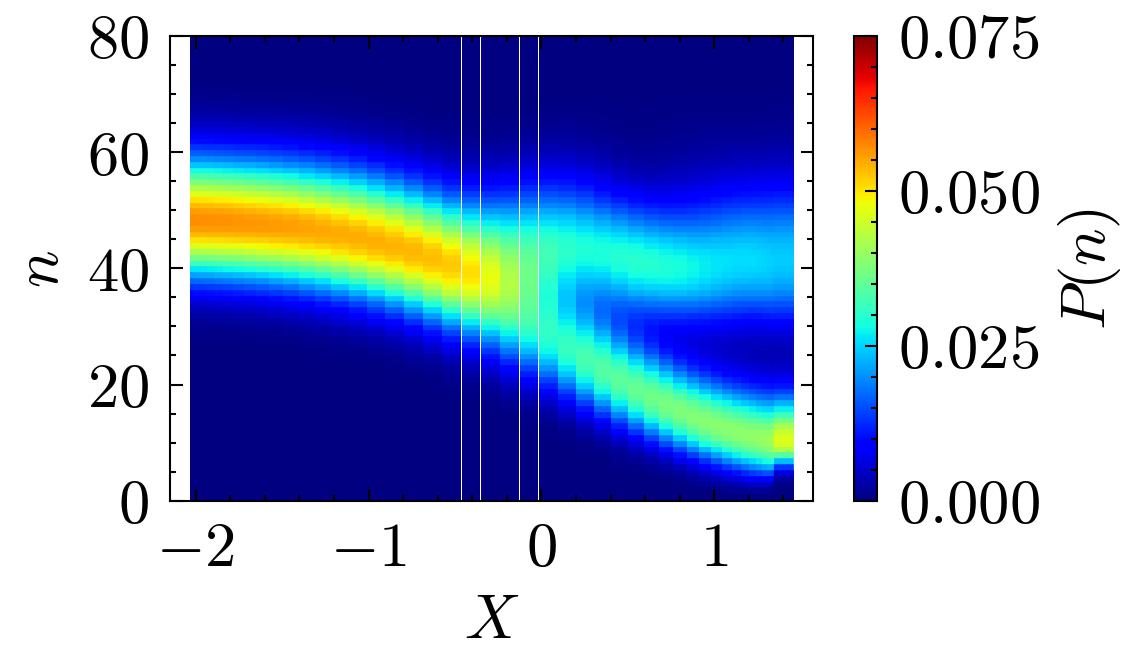} 
\includegraphics[height=3cm]{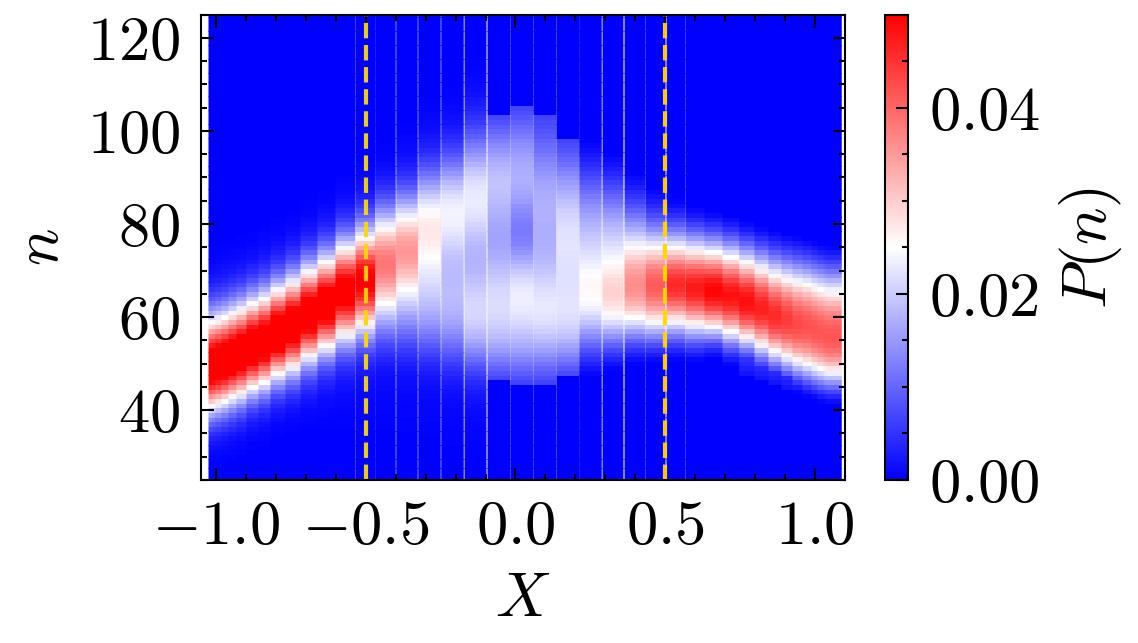} 

\caption{\justifying
{\bf Distribution vs time.}
Each column of an images is a probability distribution in a given instant of time. The time is indicated by $X=X^{qc}(t) = {\braket{X}_t}$. The upper panels show the evolution of the system energy distribution $p_{\nu}$, while in the lower panels show the evolution of the agent energy distribution $P_n$. There are 3~sets of simulations: 
{\bf (a)}~Simulation are for an $N{=}5$ dimer with ${u=3}$.  
{The sweep velocity is $v_0=0.04$.}
The agent design parameters are $N_{osc}=1000$ 
with ${\ell =0.2}$ and ${\omega=0.02}$. 
Accordingly ${\delta v_{uc}=0.004}$. \ 
{\bf (b)}~Simulation for the same dimer, 
with different initial state. 
{Here $v_0=0.2$} and ${\ell = 0.2}$ and ${\omega=0.1}$, 
and accordingly ${\delta v_{uc} =0.02}$. \
{\bf (c)}~Simulation for an $N{=}4$ trimer with ${u=1.2}$. 
It feature a sequence of two Landau-Zener transitions. 
The agent design parameters are ${N_{osc}=525}$ 
with {$v_0=0.02$} and ${\ell=0.1}$ and ${\omega=0.02}$,
such that ${\delta v_{uc}=0.002}$.
}
\label{f2}
%\end{figure*}
%
%
\vspace*{2.0cm} 
%
%
%%%%%%%%%%%%%%%%%%%%%%%%%%%%%%%%%%%%%%%%%%%%%%%%%%%%%%%
%\begin{figure*}
%\raggedright

{\bf (a)} \hspace{5cm}  {\bf (b)} \hspace{5cm}  {\bf (c)}  \\

\includegraphics[height=3.1cm]{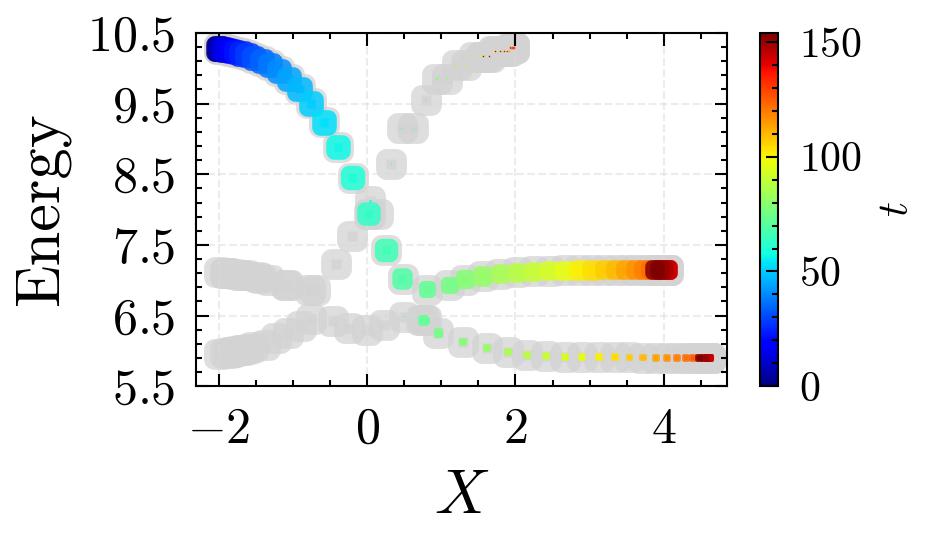}
\includegraphics[height=3.2cm]{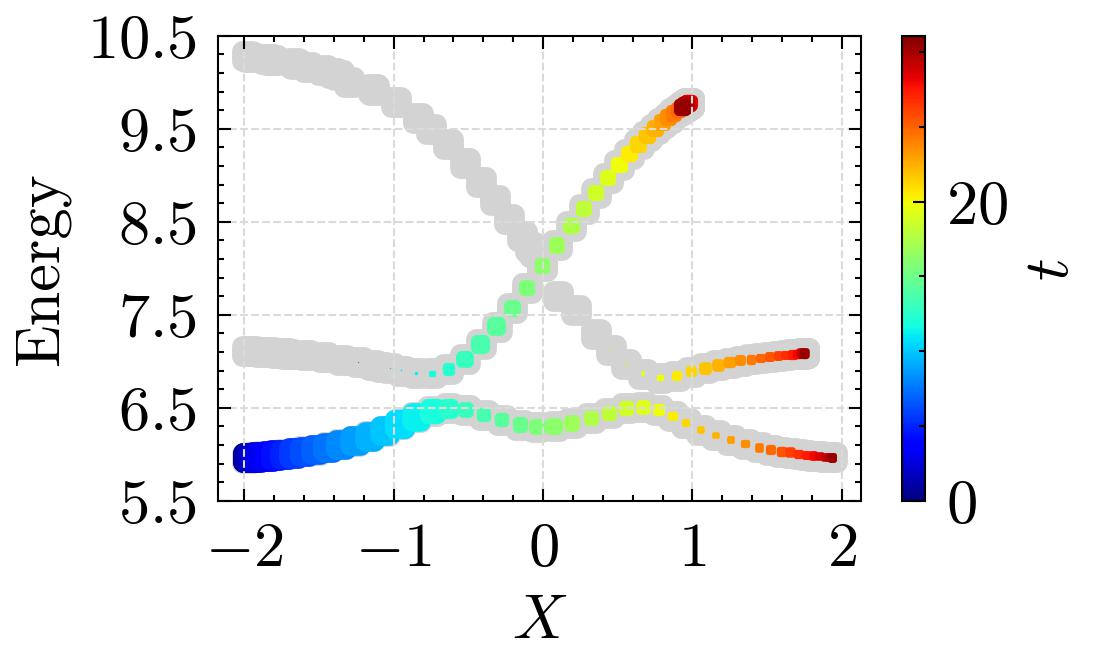} 
\includegraphics[height=3.1cm]{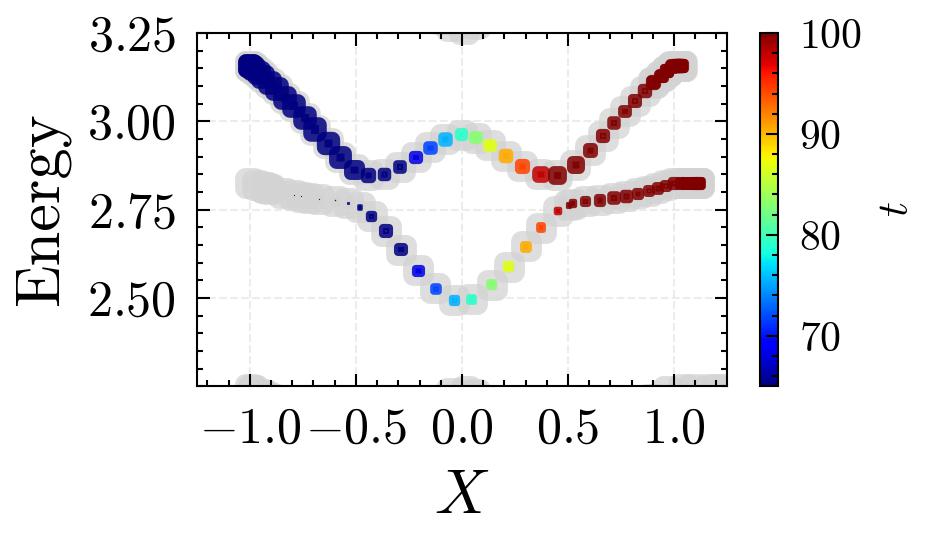} \\

\includegraphics[height=3cm]{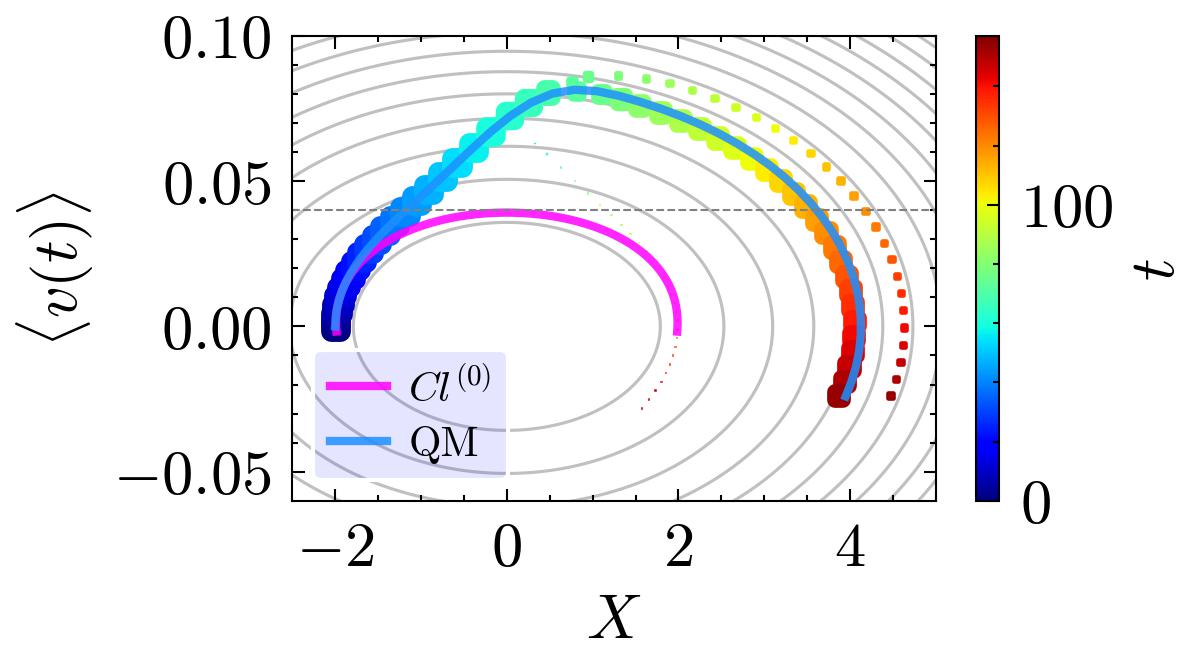} 
\includegraphics[height=3cm]{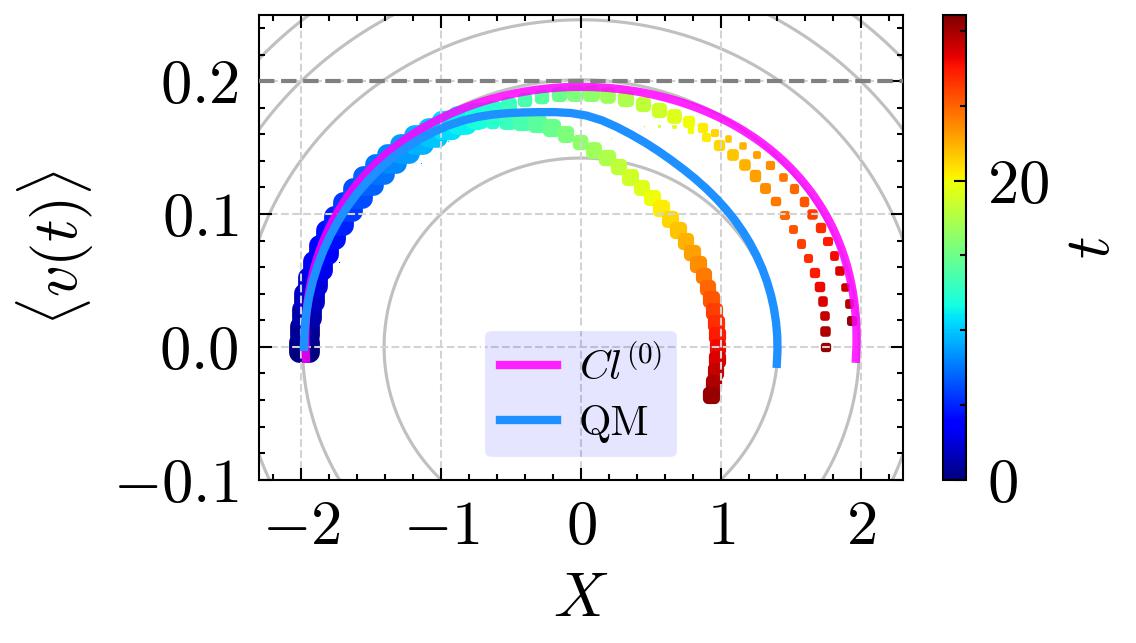} 
\includegraphics[height=3cm]{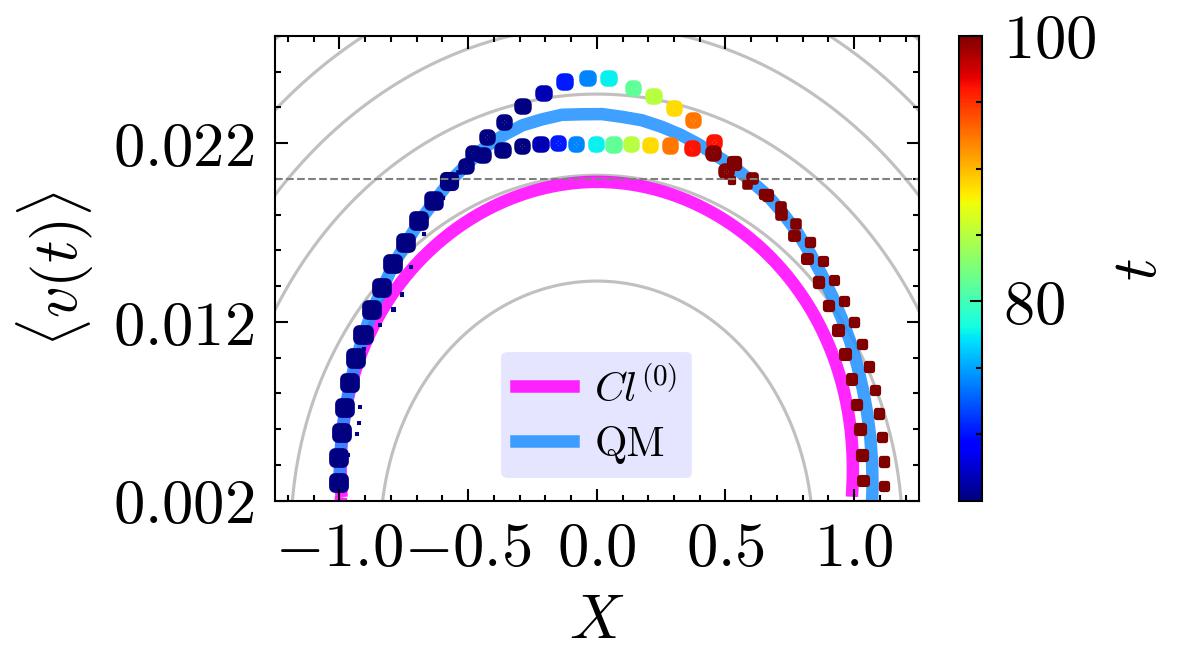} 

\caption{\justifying
{\bf Wavepacket trajectories.}
This is a more illuminating visualization of the same simulation as in \Fig{f2}. 
{\em Upper panels:} Plots of {$E_{\nu}(X)$ versus ${X=X^{\nu}(t)}$}, color-coded by time. Symbol size indicates~$p_{\nu}$. \
{\em Lower panels:} The blue line is $(\braket{X}_t,\braket{v}_t)$, 
where $v=(1/M)P$ is the agent velocity. It deviates from the classical driving (magenta line) due to back-reaction. 
The points, color coded by time, are for the evolution of each partial wave-packet.  These are the same points as in the upper panels.  
}
\label{f2traj}
\end{figure}

%%%%%%%%%%%%%%%%%%%%%%%%%%%%%%%%%%%%
\begin{figure}
%\raggedright

\centering

{\bf (a)} \hspace*{6cm} {\bf (b)} \\
\includegraphics[width=7cm]{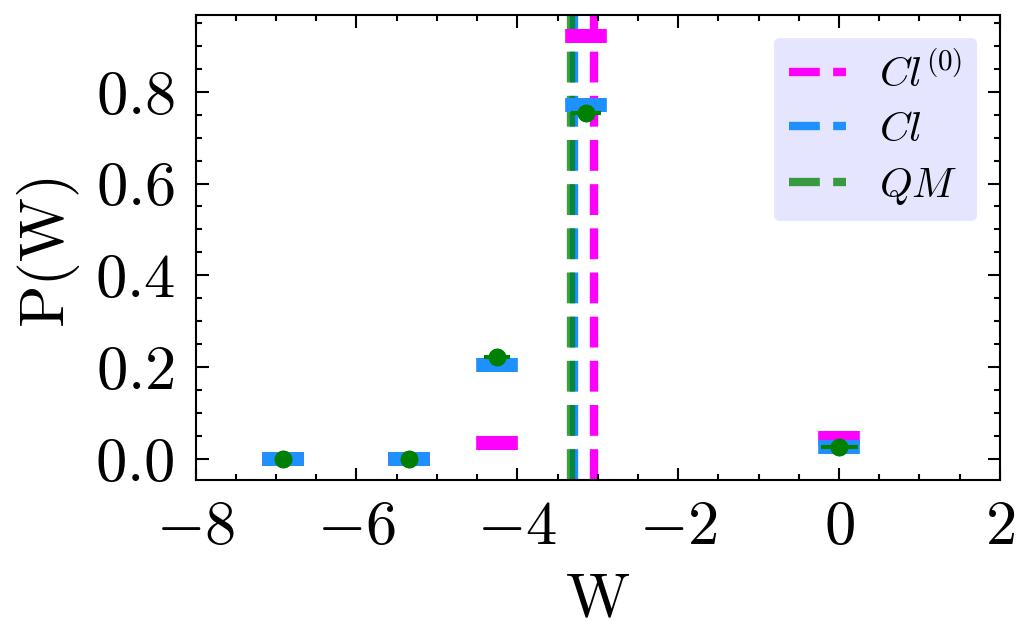}
\includegraphics[width=7cm]{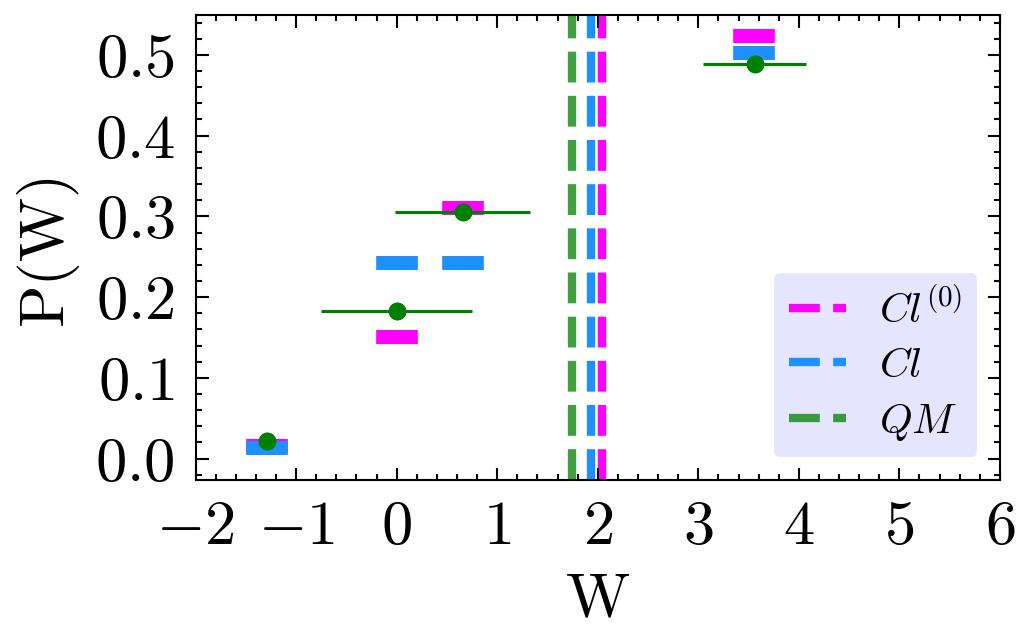}

\caption{\justifying
{\bf Final probability distribution.} 
The two panels correspond to the simulations in \Fig{f2}ab. 
Symbols indicate the probability distribution of the work done by the agent.
Vertical dashed lines indicate the average $\braket{W}$. 
Green horizontal error bars indicate the associated uncertainty in the agent energy measurement. Note that super-resolution procedure can possibly reduce this error.  
}
\label{f3}
\end{figure}

%%%%%%%%%%%%%%%%%%%%%%%%%%%%%%%%%%
\begin{figure}

\centering

\includegraphics[width=7cm]{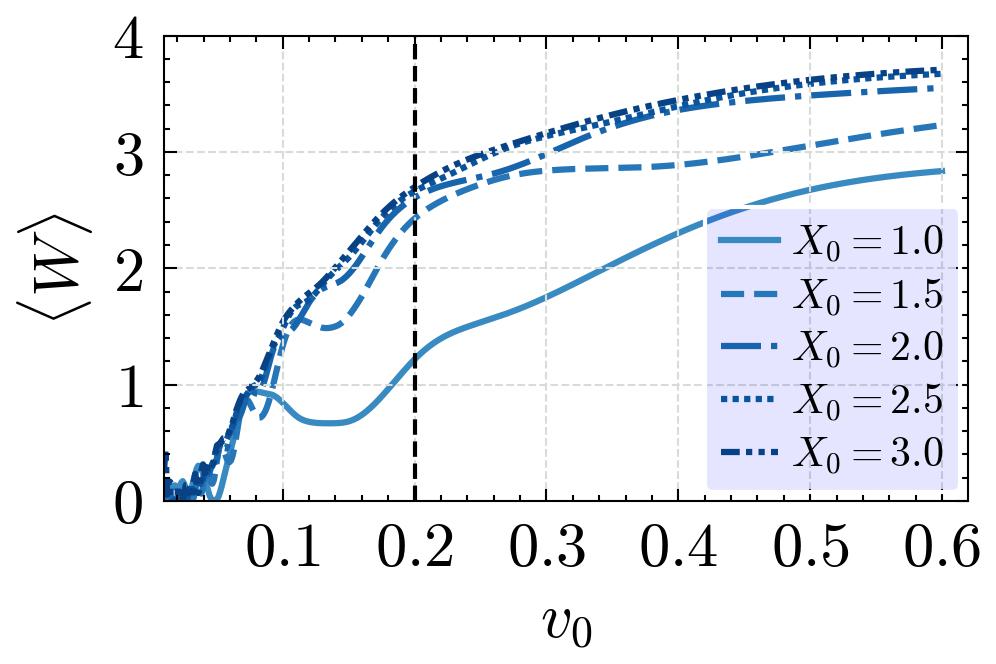}

\caption{\justifying
{\bf Average work versus sweep rate.} 
We consider here the simulation of \Fig{f2}b.
The average work is plotted versus $v_0$ 
for various sweep amplitudes $X_0$. 
The vertical line is the selected $v_0$ for the next figures.
One observes that for ${X_0 > (3/2)X_c}$ the average work is no longer sensitive to the choice of $X_0$, meaning that the agent velocity remains approximately constant during the interaction with the system.  
}
\label{f4}
%\end{figure}
%
\vspace*{13mm}
%
%%%%%%%%%%%%%%%%%%%%%%%%%%%%%%%%
%\begin{figure}
\includegraphics[width=7cm]{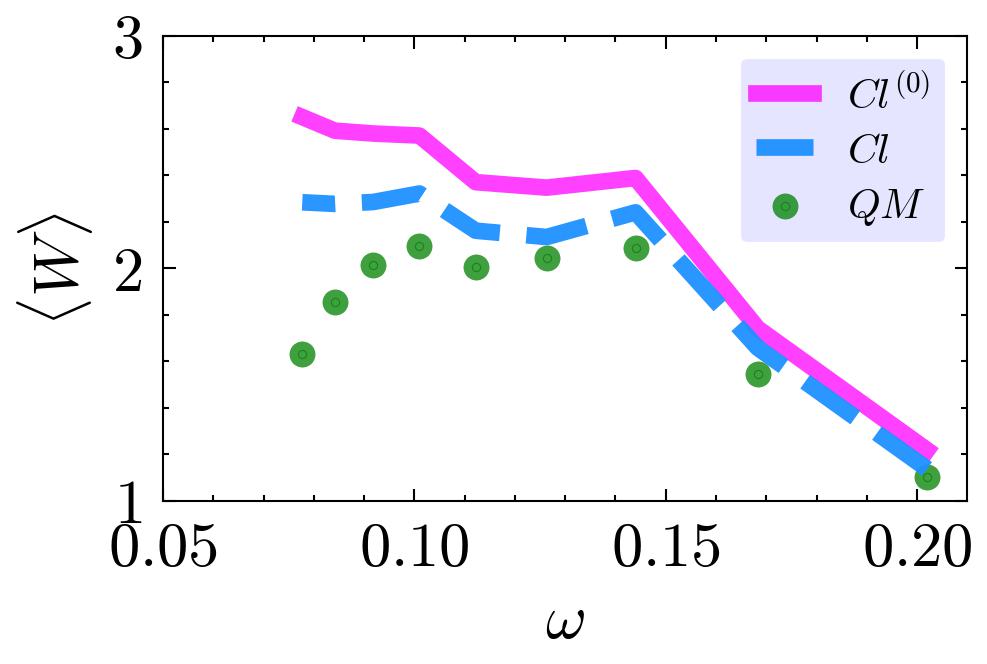}
\includegraphics[width=7cm]{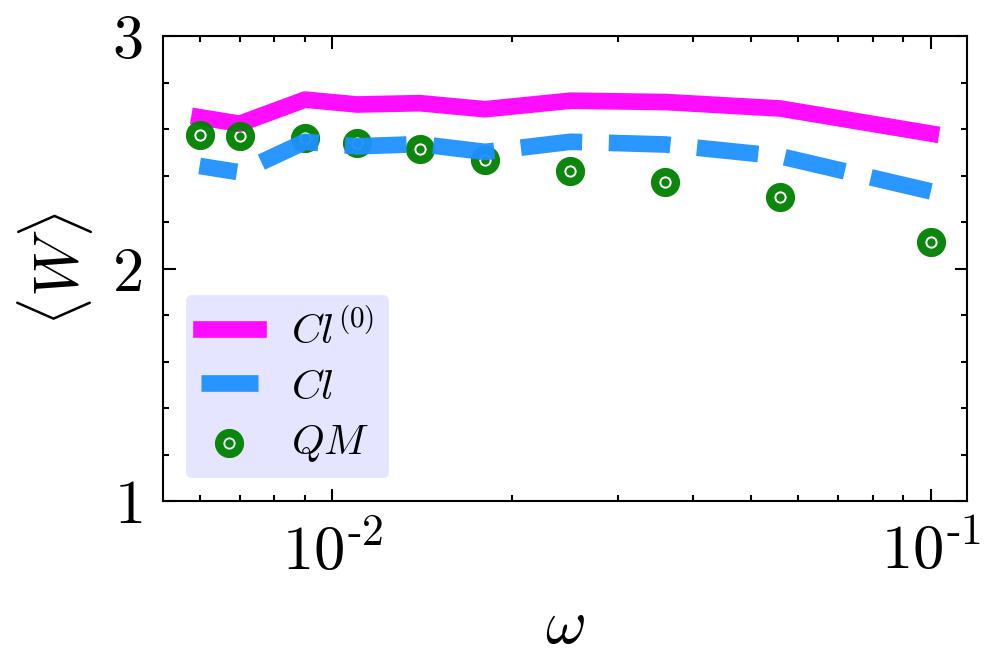}

\caption{\justifying
{\bf Dependence of $\braket{W}$ on the agent design parameters.} 
We consider here the simulation of \Fig{f2}b. 
The average work is plotted along the magenta paths of \Fig{f1}b 
as a function of $\omega$.  
The sweep amplitude $X_0$ is adjusted, 
such that the sweep rate ${v_{0}=\omega X_{0}}$ remains constant. 
The parameter $\ell$ is adjusted such that either  
${\ell^2 \omega {=} \const}$ or ${\ell \omega {=} \const}$ respectively, 
hence keeping fixed either $\delta v_{br}$ or $\delta v_{uc}$.
}
\label{f5}
\end{figure}

%%%%%%%%%%%%%%%%%%%%%%%%%%%%%%%%%%%%%%%%%%%%%%%%%%
\section{Simulations}
\label{s6s}

Three representative simulations (a,b,c) are presented in \Fig{f2}. 
{In the upper and lower panels we plot evolution of the energy probability distribution for the system ($p_{\nu}$) and for the agent ($P_n)$ respectively. The horizontal axis is the time, but it is indicated in meaningful parametric units, namely $X^{cl}(t)$. 
Yet, it is difficult to figure out what is going on. Therefore we use an improved representation of the same information in \Fig{f2traj}.     
Namely: {the horizontal axis is not  $X^{cl}(t)$ but $X^{\nu}(t)$;} only participating levels are displayed; and they are color-coded by time once populated. Thanks to that, we can follow the motion of the wavepackets in each channel, and even resolve small differences in the propagation velocities.}  

{The lower panels of \Fig{f2traj} show the variation of ${(\braket{X},\braket{v})}$ in each channel.  Without back reaction an ideal agent will move along the magenta line $X^{cl}(t)$. Due to back reaction, the trajectories $X^{\nu}(t)$ of  different channels separate. The average motion $X^{qc}(t)$ is represented by the blue line.} 

The first simulation (\Fig{f2}a), for the dimer, demonstrates the acceleration of the agent due to the variation of the system energy. Effectively there is a single LZ crossing where the probability splits between two levels. 
In the second simulation (\Fig{f2}b), the same dimer is prepared in a different level. This simulation is more interesting because it effectively involves a sequence of two LZ crossings. Each crossing splits the probability, but there is no interference.    
The third simulation (\Fig{f2}c) is for a trimer. Here effectively only two levels are populated (i.e. with non-negligible probability), and it is possible to demonstrate the effect of interference due to a sequence of two LZ crossings. 

\ \\

%%%%%%%%%%%%%%%%%%%%%%%%%%%%%%
{\em Work distribution.-- }
{The work probability distribution $P(W)$ at the end of the sweep is calculated for the full quantum mechanical simulations of \Fig{f2}. The calculation is based on the final probability distribution of the system over the levels $E_{\nu}$ via \Eq{ePcl}, or via the probabilities $P_n$ of the agent as defined in \Eq{ePW}.} 
The results for the dimer simulations (sets~a and~b) are displayed in \Fig{f3}, 
where they are labeled as $QM$. It is compared to the results with classical driving that are labeled as $Cl^{(0)}$ or as $Cl$,  depending on whether we use the $X^{cl}(t)$ of \Eq{eXcl} or the $X^{qc}(t)$ of \Eq{eXqc}, respectively. A meaningful comparison of ``quantum" to ``classical" should be with reference to $Cl$ and not with $Cl^{(0)}$.

%%%%%%%%%%%%%%%%%%%%%%%%%%%%%%
%\sect{Technical remark}
%
Few words are in order regarding the determination of the sweep amplitude~$X_0$.  A properly designed sweep should start and end outside of the interaction zone. Disregarding back-reaction, the interaction region should be fully contained in the range ${[-X_0,+X_0]}$ of the sweep. Furthermore, we want to regard ${v \sim v_0}$ as constant during the interaction.
In \Fig{f4} we test numerically the implications of changing $X_0$ with regard to the representative simulation of \Fig{f2}b. We use $\braket{W}$ as an indicator. We find that for ${X_0>(3/2)X_c}$ there is no longer sensitivity to the choice of $X_0$, indicating that the agent velocity remains approximately constant during the interaction with the system. This is important for the interpretation of the subsequent numerical results.

%%%%%%%%%%%%%%%%%%%%%%%%%%%%%%%%%%%%%%%%%%%%%%%%%%
\section{Breakdown of classicality}
\label{s7}

We use the term {\em breakdown of classicality} in order to highlight circumstances where it is not satisfactory to regard an agent as a classical entity. Let us first discuss the representative simulation whose results are displayed in \Fig{f2}a, \Fig{f2traj}a and \Fig{f3}a. In such simulation $X(t)$ can be regarded as classical coordinate. Namely, once the back-reaction is taken into account, we find that the final distribution in the QM simulation is the same as a classical simulation with ${X^{qc}(t)}$. 

Turning to \Fig{f3}b, we find that the same procedure of replacing ${X^{cl}(t)}$ by ${X^{qc}(t)}$, surprisingly, {\em spoils} the classical-to-quantum agreement. The reason becomes apparent once we inspect the temporal evolution is \Fig{f2traj}b. Most of the probability evolves diabatically to the upper level, which implies {\em average} deceleration of the agent. But this {\em average} slow-down is totally irrelevant to the second LZ crossing, because the transition in the latter is determined by the motion of the partial wavepacket at the respective channel. We therefore conclude that due to the entanglement it is not possible to define an effective classical $X(t)$ that is equally good for all the channels. 

Thus, in general, due to the entanglement, it is not possible to take into account the back-reaction within the framework of a quasi-classical description that treats $X$ as a classical coordinate. If we insist to do so, we have to minimize the back-reaction effect. Let us verify that this is indeed the case. For this purpose we change $\omega$ along the magenta horizontal segment of \Fig{f1}, away from the magenta point that indicates the representative simulation of \Fig{f2}b. The results for $\braket{W}$ are displayed in the upper panel of \Fig{f5}. We clearly see that the quantum-classical agreement is improved as the distance from the BR border is increased.       

In \Fig{f5} a second panel is displayed, where $\omega$ is changed along the other magenta segment of \Fig{f1}, that goes along a constant BR curve. As $\omega$ is increased the distance from the UC border is decreased, and as expected there is some deterioration is the classical-to-quantum agreement. On the technical side, the reason for not further increasing $\omega$ is that in order to keep $v_0$ fixed, the amplitude $X_0$ should be decreased, leading eventually to violation of our design rules.
We further discuss the significance of the uncertainty limitation in the next section, with regard to the suppression of quantum interference.

\clearpage 
%%%%%%%%%%%%%%%%%%%%%%%%%%%%%%%%%%%%%%%%%%%%%%%%%%
\section{Resolving interference}
\label{s8}

Consider a sequence of two LZ transitions. The parametric distance between them is $\delta X_0$ and the energy separation between the two interfering paths is $\delta W_0$. Let us tune the velocity $v_0$ such as to have roughly 50\%-50\% splitting. It follows that the phase accumulation is 
\beq
\delta \phi \ = \ \frac{\delta X_0}{v_0} \delta W_0  
\eeq
We assume that in an actual experiment the variation of $v_0$ allows a variation of $\delta \phi$ over several $2\pi$ periods, such that an interference pattern can be observed.
{
The condition for not having dephasing of this pattern is derived as follows: 
The travel time from the first LZ to the second LZ crossing is 
${t_{arrival} = \delta X_0/v_0}$. 
The velocity difference due to back-reaction is  
${\delta v =(1/M)[\delta W_0/v_0] }$.
The difference in the arrival time is 
${\delta t_{arrival} = (\delta X_0/v_0^2) \delta v  }$. 
The travel time of the finite-width wavepacket via the second LZ crossing is 
${ \delta t_{uc} = \ell / v_0 }$.
The condition for not having dephasing is 
${\delta t_{arrival} < \delta t_{uc}}$.
}
Thus we get the condition 
\beq \label{eVintrf}
\delta v_{uc} \ \ < \ \ \frac{v_0^2}{\delta X_0 \delta W_0} 
\eeq
{This condition has the form ${\delta v_{uc}<\delta v_0}$ as \Eq{eUC} with} 
\beq \label{edv0}
\delta v_0 \ = \ \frac{1}{\delta \phi}v_0
\eeq
%
%where the $2\pi$ is inserted for stylistic reason.
%
We thus identified what is the $\delta v_0$ that is required for the purpose of resolving interference in work measurement.  It is important to realize that the condition here is on $\delta v_{uc}$ and not on $\delta v_{br}$, reflecting that we are dealing here with a quantum condition.

%%%%%%%%%%%%%%%%%%%%%%%%%%%%%%%%%%%%%%%%%%%%%%%%%
\begin{figure}[t!]
\centering

\includegraphics[width=10cm]{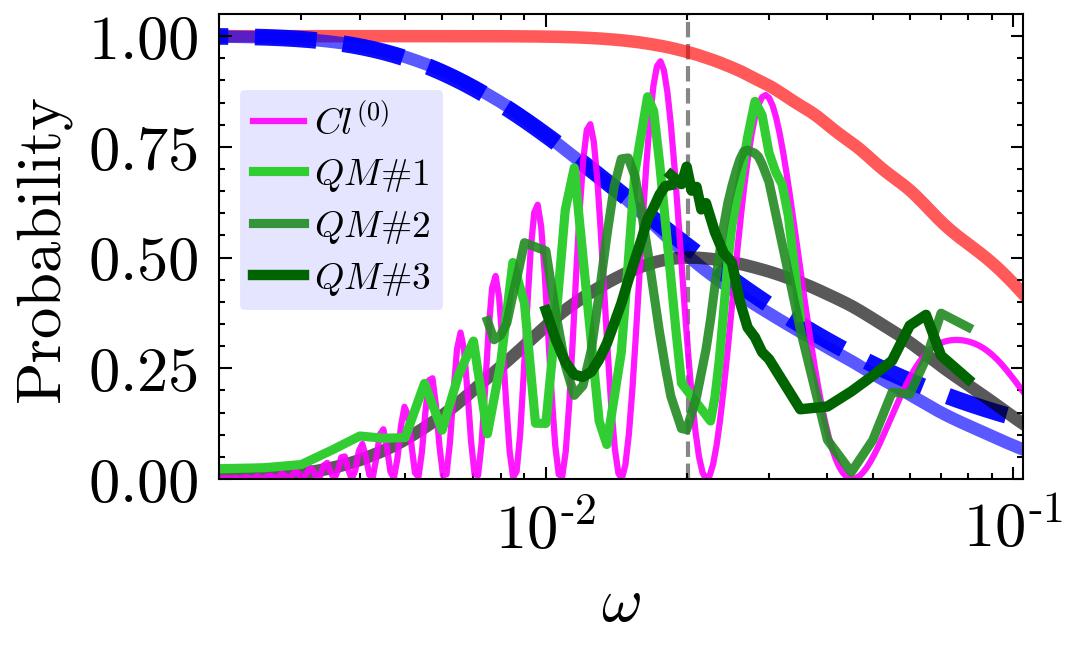} 

\includegraphics[width=10cm]{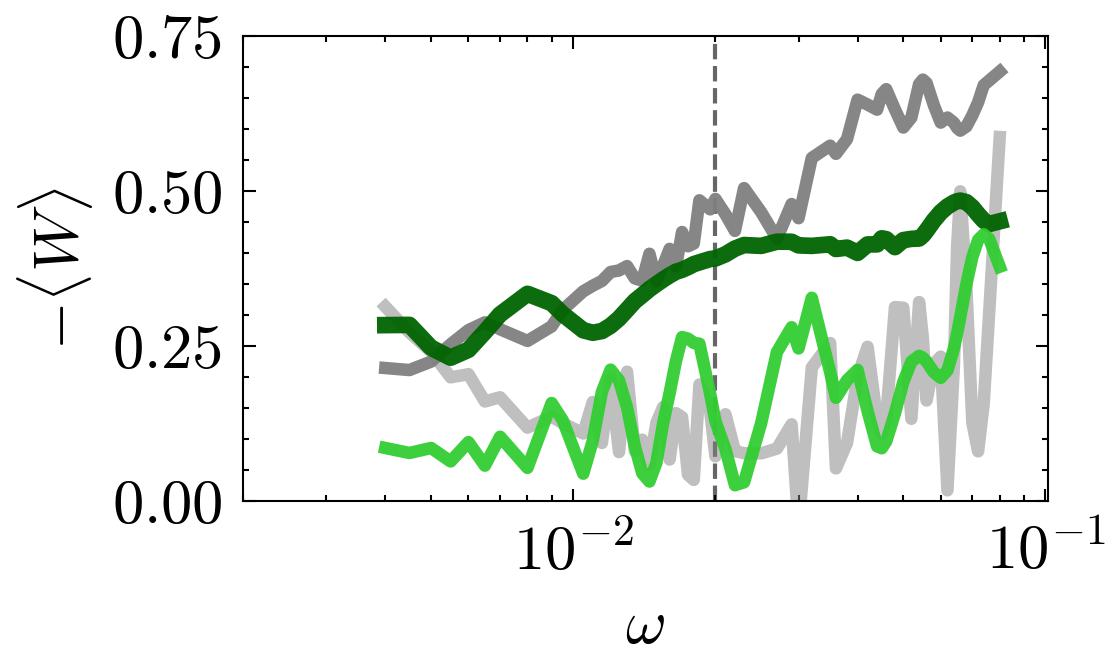} 

\caption{\justifying
{\bf Effect of interference.} 
{\bf (a)}~{\em Upper panel:} We consider simulations as in \Fig{f2}c, with different values of $\omega$. Red curve is the survival probability in the two adiabatic energy levels that are plotted there. Blue solid curve is the probability ($1{-}P_{LZ}$) to stay in the upper level after the first LZ crossing, compared with the analytical LZ formula (dashed line).  From this curve we deduce that the velocity for getting 50\%-50\% splitting is ${v_0=0.0212}$, 
which is indicated by vertical dashed line.
Magenta curve is the probability to end up in the lower level at the end of a classical sweep. The sweep amplitude is ${X_0=1.0606}$. The results of simulations with a quantum agent are plotted with shades of green lines for ${\ell =0.05,0.10,0.20}$. As $\ell$ is increased the interference is diminished. The Gray line indicates the expected result for total dephasing.
{{\bf (b)}~{\em Lower panel:} Average work done. The light and dark green lines refer to the same simulations of the upper panel that were labeled as QM\#1 and QM\#3 respectively. The additional light and the dark gray lines refer to simulations whose duration has been doubled, namely, ${t=2\pi/\omega}$ instead of ${t=\pi/\omega}$.}
}
\label{f6}
\end{figure}

Referring to the simulation of \Fig{f2}c we have verified in \Fig{f6}a that the non-adiabatic transition probability in the first LZ crossing fits well the LZ formula 
${P_{\text{LZ}}=\exp(-\const/\dot{X}^2) }$ with ${\dot{X}:=X_0\omega}$. The probability to stay in the initial adiabatic level after the first transition is $1{-}P_{\text{LZ}}$, {see blue line in \Fig{f6}a}.  {For a classical driving the probability to end up in the lower level at the end of a classical sweep is given by the Magenta curve.}
Assuming that the second transition features the same transition probability as the first transition, one deduces that the final probability to stay in the initial adiabatic level can be approximated by the formula 
\beq
p \ \ = \ \ \left|(1{-}P_{\text{LZ}}) + P_{\text{LZ}}e^{i \delta \phi}\right|^2 
% = 1 - 4(1-q)q sin^2(phase/2)
\eeq
If we had total dephasing the result would be 
\beq
p \ \ = \ \ P_{\text{LZ}}^2+(1{-}P_{\text{LZ}})^2
\eeq
as demonstrated by in the figure by the gray line.

In the agent-driven case we have partial dephasing that depends on $\ell$. This is demonstrated in \Fig{f6}a by the green lines. This probability is a feature of $P(W)$. Optionally one can calculate $\braket{W}$, as in \Fig{f6}b, or other moments of the distribution. We observe that the quantum signature of  interference is controlled by the small parameter $\delta v_{uc}/\delta v_0$ with \Eq{edv0}. Extra technical details regarding the design of the simulation are provided in \App{sDetails}. 

{The observed dephasing effect is possibly not impressive in the demonstration of \Fig{f6}, but for sweep that involves many interference loops the suppression effect will accumulate. Using different phrasing: the quantum fluctuations are signature of quantum interference and their suppression is expected to depend on the small parameter that has been deduced above. Numerically it is too demanding to demonstrate this statement for a multi-crossing sweep, but for illustration purpose we can simply double the time duration of the simulation. The outcome are the gray curves in \Fig{f6}b. The observed fluctuations are the fingerprint of interference, and the $\ell$~dependence is apparent.}

{We note that a parallel study \cite{nfx} has considered suppression of interference in a periodically agent-driven {\em two-level system} that features a {\em single} LZ crossing per cycle. There, the size of the interference loop was determined by the agent amplitude $X_0$, and not by an intrinsic parametric scale $\delta X_0$. Accordingly, the discussion there is {\em irrelevant} to the agent design problem as formulated in the present work. Nevertheless, the suppression of interference for a {\em multi}-cycle process has been nicely demonstrated there (Fig.5).}

%%%%%%%%%%%%%%%%%%%%%%%%%%%%%%%%%%%%%%%%%%%%%%%%%%
%\newpage 

\clearpage
%%%%%%%%%%%%%%%%%%%%%%%%%%%%%%%%%%%%%%%%%%%%%%%%%%
\section{Summary and Discussion}
\label{s10}

It is common to regard baths and work agents as classical entities. But in a mesoscopic reality one wonders what are the implications of quantum mechanics. While quantum dissipation (with baths) is a rather well developed arena, little attention has been paid for the work agent. The crucial question is whether it can be regarded as a classical coordinate $X(t)$ that merely experiences some energy variation due to back-reaction. Such perspective neglects two major aspects of quantum mechanics: the quantum uncertainty $\ell$ of any dynamical coordinate, and its multi-channel entanglement with the driven system. 

To elucidate the rationale underlying the present study, it is instructive to consider an analogy. In Quantum Mechanics there is a free parameter $\hbar$ that governs the emergence of quantum effects; in the limit $\hbar \rightarrow 0$ the theory reduces to its classical counterpart. Also in our discussion of work-agent, we have free {\em design} parameters. Clearly they affect the outcome of the measurement. But we say that there is a limit of ``classical infinite mass agent"  where the outcome $P(W)$ becomes design-independent. This is the idealized limit where e.g. the Jarzynski equality becomes applicable. Our purpose was to provide the  conditions for reaching this idealized regime.         

In this study we have explored the major reasons for the breakdown of classicality. The agent cannot be regarded as a classical entity because it is impossible in general to define a single $X^{qc}(t)$ that replaces the separate $X^{\nu}(t)$ of the entangled channels. This problem can be minimized if we take a distance from the BR border in the design diagram \Fig{f1}. 

The other reason for the breakdown of classicality is related to the quantum uncertainty $\ell$.  In a previous work \cite{nfa} the concern was that small $\ell$ (large $n_{ph}$) may blur the probability distribution $P(W)$. This is true, but one may argue that maybe it can be addressed by super-resolution procedure. 
{The present work argues that irrespective of this resolution issue, the uncertainty $\ell$ also affects the feasibility to detect quantum interference if the condition \Eq{eVintrf} is broken. This condition is expressed in terms of parametric scales $\delta X_0$ and $\delta W_0$ that characterize the spectral properties of the driven system. It is a quantum condition that has no analogue in the classical back-reaction analysis.}

{Several existing experiments are close to the regime considered here. We already pointed out in subsection~1.1 that Han, Katz, and Sela~\cite{nfx} have provided the most direct setting, where an LC mode is the dynamical work agent.
Kervinen et al.~\cite{Kervinen2019} and Du, Lan, and Yu~\cite{DuLanYu2013} show that Landau-Zener-Stueckelberg interference can be measured in superconducting and hybrid qubit systems. These experiments are relevant to our interference condition, because they realize repeated avoided-crossing transitions with measurable interference visibility.
Atalaya et al.~\cite{Atalaya2024} show that very small coherent or thermal photon populations in microwave resonators can be resolved experimentally. This is relevant to the readout side of the work-agent protocol, where work is inferred from the energy change of the oscillator.
Pekola et al.~\cite{Qmeas} provide a related superconducting calorimetric approach, where energy exchange is measured at the single-quantum level. This supports the broader thermodynamic idea of monitoring an external degree of freedom, although in the present work this degree of freedom is a coherent work agent rather than a passive calorimeter.} 

%Saira et al.~\cite{saira2012test}, Koski et al.~\cite{koski2013distribution}, and Barker et al.~\cite{barker2022experimental} show that dissipated energy, entropy production, and work fluctuations can be measured in mesoscopic electronic devices. In those experiments the gate is usually treated as a prescribed classical drive. Our work asks when such a drive can instead be treated as a finite quantum work agent, and when its back-reaction and quantum uncertainty become observable.}

Finally, we note that the implementation of Ramsey-like spectroscopic methods become more feasible if they are intended to measure {\em work} in the strict sense, namely, changes in the energy content of a relatively simple work-agent that can be engineered for this stated purpose.    

\ \\ \ \\ 
%%%%%%%%%%%%%%%%%%%%%%%%%%%%%%%%%%%%%%
{\bf Acknowledgments} --  
We thank Eran Sela for fruitful discussions. The research has been supported by the Israel Science Foundation, grant No.518/22.

%%%%%%%%%%%%%%%%%%%%%%%%%%%
\newpage 
\appendix

%%%%%%%%%%%%%%%%%%%%%%%%%%%%%%%%%%%%%%%%%%%%%%%%%%
\section{The ideal agent}
\label{sIdeal}

Here we prove that an $M{=}\infty$ quantum agent is equivalent to a classical agent. Consider a piston of mass $M$ and expand its kinetic energy as 
${E = [1/(2M)]P_0^2 + v_0 (P{-}P_0) + [1/(2M)](P{-}P_0)^2}$, where ${v_0=(1/M)P_0}$. 
In the $M{=}\infty$ limit, dropping a constant, we get the idealized dispersion ${E = v_0  P_0 }$.  We now prove quantum-to-classical equivalence in this limit. We write the Hamiltonian schematically as ${H_{total}(r,p;X,P)=H(r,p;X)+v_0 P}$, and use the shorter notation ${H(X)\equiv H(r,p;X)}$. The Feynman path-integral expression for the propagator of the classically-driven system is 
\beq \nonumber
&& U(r|r_0; U(r|r_0; X^{cl})) =
\\  \label{eUcl}
&& \int d[r] \prod_j \BraKet{r_{j{+}1}}{e^{-i\delta t H(X^{cl}(t_j))}}{r_j} 
\eeq
where 
\beq
X^{cl}(t')=v_0 t', 
\ \ \ \ \ \  
t'\in [0,t]
\eeq
It is implicitly assumed that the initial time ${t'=0}$ and the final time ${t'=t}$ are outside of the scattering region.  
Next we write the path-integral expression for the quantum coupled system
\beq \nonumber
&&U(r,X|r_0,X_0) 
\\ \nonumber
&& =\iint d[r]dX \prod_j 
\BraKet{r_{j{+}1},X_{j{+}1}}{e^{-i\delta t H(X)+v_0P}}{r_j,X_j} 
\\ \nonumber
&& =\int d[r] \prod_j 
\BraKet{r_{j{+}1}}{e^{-i\delta t H(X_j)}}{r_j} \delta(X_{j{+}1}{-}X_j{-}v_0\delta t ) 
\\ \nonumber
&& = U(r|r_0; X_0+X^{cl}) \ \delta(X - X_0 - v_0 t) 
\eeq
The quantum propagator has factorized, and has reduced to that of a classically-driven system. It contains a time-shifted version of \Eq{eUcl}, namely, 
\beq \nonumber
&& U(r|r_0; X_0+X^{cl}) = 
\\
&& e^{i \frac{X_0}{v_0} H(X^{cl}(t))} \ U(r|r_0;X^{cl}) \ e^{-i\frac{X_0}{v_0}H(X^{cl}(0))}  
\ \ \ 
\eeq
The operation of $U(r,X|r_0,X_0)$ on a wavepacket 
leads to a distance $v_0 t$ displacement of the agent. 
The change in its momentum depends on the work.  
Namely, it is 
\beq
\delta P = -\frac{1}{v_0} \left[ E_{\nu}(X^{cl}(t)) - E_{\nu_0}(X^{cl}(0)) \right] 
\eeq
for a transition from the initial channel $\nu_0$ to a final channel $\nu$.

%%%%%%%%%%%%%%%%%%%%%%%%%%%%%%%%%%%%%%%%%%%%%%%%%%
\section{Measurement of Energy}
\label{s9}

For completeness, we briefly review how changes in energy can be measured. The naive approach for obtaining the probability distribution of $E_B-E_A$ is the {\em two point measurement scheme}. While such a procedure is feasible for microscopic systems (e.g., a single atom), it becomes problematic for many-body mesoscopic systems where energy is not a simple observable, rendering this approach seemingly impractical. Nevertheless, this difficulty can be dealt, in principle, by implementing a Ramsey-type spectroscopic protocol \cite{fusco2014assessing,ancilla1,ancilla2,ancilla4}. 
In this method, it is the generating function $F(\tau)$ that is measured, aka fidelity amplitude. 
It is defined as the Fourier transform of $P(W)$. From \Eq{ePcl}, disregarding sign convention for $W$, we get  
\beq  \nonumber
F(\tau) &=& \BraKet{\Psi}{U^{\dagger} e^{i\tau H^{(B)}} U e^{-i\tau H^{(A)}} }{\Psi} \\
&=& \Braket{ \ U_B U \,\Psi \ }{ \ U U_A \, \Psi \ }    
\eeq
where ${ \ket{\Psi} = \ket{\nu_0} }$ is the initial state of the system, 
and ${U_{A,B} = \exp[-i\tau H^{(A,B)}]}$ is the free system evolution
with the Hamiltonians of \Eq{eHAB}.

\begin{figure}
%\raggedright

\centering

\ \ \ \ \ {\bf (a)} \\
\includegraphics[width=9cm]{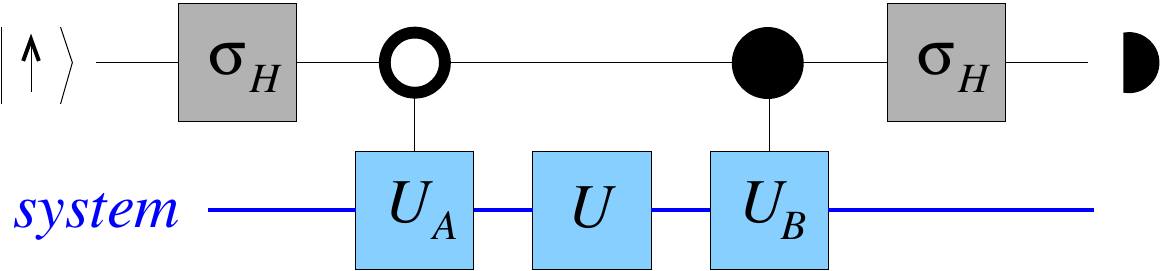} 

\ \\ \ \\ 

\ \ \ \ \ {\bf (b)} \\
\includegraphics[width=9cm]{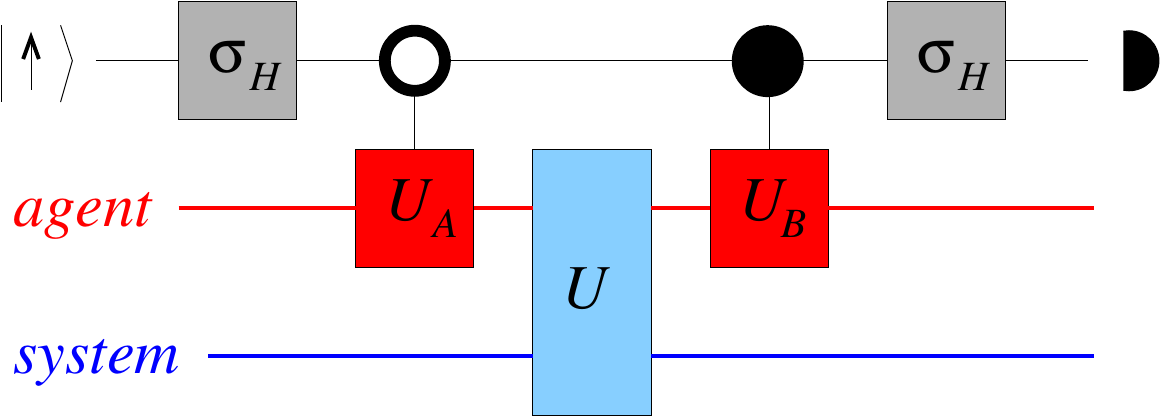} 

\caption{\justifying
{\bf Measurement circuit.} 
{The upper panel illustrates a circuit that is intended for measurements that characterize the statistics of changes in the energy of the system.} Notations are as in Fig.1 of \cite{ancilla1}. The system  is prepared in channel $\nu_0$. More generally it might be in a thermal state. An ancilla qubit is incorporated. The $\sigma_H$ indicates Hadamard operation. The probability to find the ancilla in the $\uparrow$ state is measured, and from that the fidelity amplitude $F(\tau)$ is extracted.  
The circuit in the lower panel is a variation on the same idea. Here the purpose is to characterize the statistics of the work that is performed by an agent. The agent is prepared in a coherent state $\alpha_0$, with $X{\sim}-X_0$ and $P{\sim}0$.    
}
\label{f7}
\end{figure}

The Ramsey measurement procedure is illustrated in \Fig{f7}a. 
It requires the incorporation of an ancilla qubit. 
The final state of the circuit is 
\beq 
\ket{\text{Circuit}} = \frac{1}{\sqrt{2}} \Big[ \ket{\uparrow} \otimes UU_A \ket{\Psi} + \ket{\downarrow}\otimes {U_B}U \ket{\Psi} \Big] 
\ \ \ \  
\eeq
Hence the probability to find the ancilla in the $\uparrow$ state is 
\beq
\text{Probability}  \ = \  \frac{1}{2} \Big\{1+\re[F(\tau) e^{i\varphi}] \Big\} 
\eeq
where the interferometric phase $\varphi$ is due to optional detuning of the ancilla (not indicated explicitly in \Fig{f7}).

{We want to measure {\em work}, which is the change in the energy of the agent, not of the system. 
A simple two point measurement scheme would be the natural procedure to adapt, because the agent is a simple degree of freedom. Nevertheless we discuss below the optional adaptation of the Ramsey-type method.}

%%%%%%%%%%%%%%%%%%%%%%%%%%%%%%%%%%%%%%%%%%%%%%%%%

Referring to \Eq{ePw}, we can use an analogous definition for $F(\tau)$ that involves ${ \ket{\Psi} = \ket{n_0,\nu_0} }$, and ${H^{(A)}=H^{(B)}=H_{agent}}$, as illustrated in \Fig{f7}b. The problem is that we do not want to start with an $n$ state. We want to consider classical-like driving scenario for which the initial state is a coherent state. An ad hoc definition of $F(\tau)$ with initial state ${ \ket{\Psi} = \ket{\alpha_0,\nu_0} }$ will give a  Kirkwood-Dirac quasiprobability distribution \cite{KDQ} which is apparently not useful in the present context. 
Nevertheless, there is a simple way to overcome the formal difficulty. We can define $H^{(A)}$ as the {\em preparation Hamiltonian} of the agent \Eq{eHPrep}. Thus the proposed setup is
\beq
\ket{\Psi} = \ket{\alpha_0,\nu_0}, 
\hspace{2cm}
H^{(A)}=H_{agent}^{(A)},
\hspace{2cm}
H^{(B)}=H_{agent}
\eeq
By default the ideal preparation would be with ${n_0^{(A)}=0}$, but all the formulas above are straightforwardly generalized also for thermal mixtures of energy-eigenstates.    

{A final clarification is in order:} The scheme of \Fig{f7}b is somewhat misleading. In practice we assume that we can turn on and off parameters that control the agent, e.g. as in \cite{FolmanWork}. 
In fact we assume that during the time $\tau$ the agent is either ``on" or ``off" as dictated by the ancilla. But we cannot turn off the evolution of the system. This means that if the agent is off, we still have ${H_{total} = H + 0 \cdot H_{agent}}$. 
It follows that the proper interpretation of the evolution operator is 
${U=\exp(-i\tau  H)\exp(-it  H_{total})\exp(-i\tau  H) }$. This does not introduce any complication as far as the spectroscopy is concerned: the extra pre-evolution and post-evolution stages have no implication on the work, because they take place outside of the scattering region. Namely, the sweep setup is designed such that 
${U=\exp(-i\tau  H^{(B)})\exp(-it  H_{total})\exp(-i\tau  H^{(A)}) }$.

%%%%%%%%%%%%%%%%%%%%%%%%%%%%%%%%%%%%%%%%%%%%%%% 
\section{{Change of basis}}
\label{sBasis}

The standard basis is the occupation basis:  $\ket{n_{osc},n_{sites}}$, 
where $n_{site}$ is an index that distinguishes the BHH Fock states.  
The wavefunction is represented by  $\Psi_{n_{osc},n_{sites}}$. 
There are two transformation matrices that are determined respectively via diagonalization of the $X$ operator and of the $H(sys,X)$ operator:
\beq
T_{n_{osc},X} &=& \Braket{n_{osc}}{X}   
\\
T(X)_{n_{sites},\nu}  &=&  \Braket{n_{sites}}{\nu(X)}
\eeq
The change of basis can be done in two steps:
\beq
\Psi_{X,n_{sites}} &=& \sum_{n_{osc}} [T^{\dagger}]_{X,n_{osc}}  \Psi_{n_{osc},n_{sites}} 
\\
\Psi_{X,\nu}  &=& \sum_{n_{sites}}  [T(X)^{\dagger}]_{\nu,n_{sites}}  \Psi_{X,n_{sites}} 
\eeq
An approximation is to use a single T-matrix that is calculated for  $X=\braket{X}$, namely, 
\beq
\Psi_{X,\nu}  \approx \sum_{n_{sites}}  [T(\braket{X})^{\dagger}]_{\nu,n_{sites}}  \Psi_{X,n_{sites}} 
\eeq

%%%%%%%%%%%%%%%%%%%%%%%%%%%%%%%%%%%%%%%%%%%
\section{Details regarding simulation} 
\label{sDetails}

Providing some extra details regarding the simulations of \Fig{f6} is appropriate. We are dealing with an ${N=4}$ trimer coupled to a truncated oscillator that has ${N_{osc}=525}$ levels. The interaction range is ${X_{c} = 1}$. The chosen sweep amplitude ${X_{0} = 1.0606}$ was verified as appropriate. The optimal velocity for 50\% splitting at each LZ crossing was determined numerically, and it is ${v_{0} = \omega X_{0} = 0.021}$. 

The parameters that characterize the interference loop are by inspection ${\delta X_{0} \sim 1}$, and ${\delta W_{0} \sim 1/2}$, for which the phase accumulation is ${\delta \phi \sim 24}$. The desired velocity resolution by \Eq{eVintrf} is ${\delta v_0 = 0.00089}$.

The simulations were carried out for ${\ell = 0.05, 0.10,0.2}$.  
Respectively, the uncertainty in velocity was 
${\delta v_{uc} \sim 0.001,0.002,0.003}$, 
while the back reaction was 
${\delta v_{br} \sim 0.001,0.005,0.019}$. 
Note that for this design 
$\delta v_{uc}/\delta v_{0}$ is larger than unity, 
while  $\delta v_{br}/v_{0}$ is smaller than unity. 

%$\dfrac{\delta v_{uc}}{\delta v_{0}} = 1.12,2.24,3.36 $ ; \ \ \ 
%$\dfrac{\delta v_{br}}{v_{0}} =0.055,0.22,0.89 $ \\

\ \\ \ \\ 

%\clearpage 

%%%%%%%%%%%%%%%%%%%%%%%%%%%%%%%%%%%%%%%%
%\section*{References}

%\bibliography{references}

%%%%%%%%%%%%%%%%%%%%%%%%%%%%%%%%%%%%%%%
%

%%%%%%%%%%%%%%%%%%%%%%%%%%%%%%%%%%%%%%%%%%%%%%%%%%%%%%%%%%%%%%
%%%%%%%%%%%%%%%%%%%%%%%%%%%%%%%%%%%%%%%%%%%%%%%%%%%%%%%%%%%%%%
\end{document}